\documentclass{elsart}
\usepackage{amssymb}
\usepackage{amsmath}
\usepackage{braket}
\usepackage[utf8]{inputenc}
\usepackage{subfig}
\usepackage[pdftex]{graphicx}
\usepackage{color}
\usepackage{url}

\newcommand{\tit}[1]{\textit{#1}}

\newcommand{\ttt}[1]{\texttt{#1}}

\newcommand{\eff}{\mathrm{eff}}
\newcommand{\mrm}[1]{\mathrm{#1}}
\renewcommand{\min}{\mathrm{min}}
\renewcommand{\max}{\mathrm{max}}
\newcommand{\inrm}{\mathrm{in}}
\newcommand{\outrm}{\mathrm{out}}
\newcommand{\ctp}{\mathrm{ctp}}

\begin{document}

\begin{frontmatter}

\title{\ttt{dftatom}: A robust and general Schr\"odinger and Dirac solver for
atomic structure calculations}

\author[fzu,unr,mff]{Ond\v rej \v Cert\' ik},
\ead{ondrej.certik@gmail.com}
\author[llnl]{John E. Pask},
\ead{pask1@llnl.gov}
\author[fzu]{Ji\v r\'i Vack\'a\v r}
\ead{vackar@fzu.cz}

\address[fzu]{Institute of Physics, Academy of Sciences of the Czech
Republic, Na Slovance 2, 182 21 Praha 8, Czech Republic}
\address[unr]{University of Nevada, Reno - 1664 N. Virginia St., Reno, NV
89557-0208, USA}
\address[mff]{Faculty of Mathematics and Physics, Charles University, Ke
Karlovu 3, 121 16 Praha 2, Czech Republic}
\address[llnl]{Lawrence Livermore National Laboratory, 7000 East Avenue, Livermore, CA 94550, USA}

\begin{abstract}
A robust and general solver for the radial Schr\"odinger, Dirac, and Kohn--Sham
equations is presented. The formulation admits general potentials and meshes:
uniform, exponential, or other defined by nodal distribution and derivative
functions. For a given mesh type, convergence can be controlled systematically
by increasing the number of grid points. Radial integrations are carried out
using a combination of asymptotic forms, Runge-Kutta, and implicit Adams
methods. Eigenfunctions are determined by a combination of bisection and
perturbation methods for robustness and speed. An outward Poisson
integration is employed to increase accuracy in the core region, allowing
absolute accuracies of $10^{-8}$ Hartree to be attained for total energies of
heavy atoms such as uranium. Detailed convergence studies are presented and
computational parameters are provided to achieve accuracies commonly required
in practice. Comparisons to analytic and current-benchmark density-functional
results for atomic number $Z$ = 1--92 are presented, verifying and providing a
refinement to current benchmarks. An efficient, modular Fortran 95
implementation, \ttt{dftatom}, is provided as open source, including
examples, tests, and wrappers for interface to other languages; wherein
particular emphasis is placed on the independence (no global variables),
reusability, and generality of the individual routines.
\end{abstract}

\end{frontmatter}

{\bf PROGRAM SUMMARY}

\begin{small}
\noindent
{\em Manuscript Title:} \ttt{dftatom}: A robust and general Schr\"odinger and Dirac solver for atomic structure calulations  \\
{\em Authors:}  Ond\v rej \v Cert\' ik, John E. Pask, Ji\v r\'i Vack\'a\v r  \\
{\em Program Title:}  dftatom                                  \\
{\em Journal Reference:}                                      \\
{\em Catalogue identifier:}                                   \\
{\em Licensing provisions:}  MIT license                        \\
{\em Programming language:}  Fortran 95 with interfaces to Python and C \\
{\em Computer:}  Any computer with Fortran 95 compiler         \\
{\em Operating system:}  Any OS with Fortran 95 compiler \\
{\em RAM:}  500 MB                                              \\
{\em Number of processors used:}                              
{\em Classification:}  2.1 Structure and Properties               \\
{\em Nature of problem:} \\
Solution of the Schr\"odinger, Dirac, and Kohn--Sham equations of Density Functional Theory for isolated atoms. \\
{\em Solution method:} \\
Radial integrations are carried out using a combination of asymptotic forms, Runge-Kutta, and implicit Adams methods. Eigenfunctions are determined by a combination of bisection and perturbation methods. An outward Poisson integration is employed to increase accuracy in the core region. Self-consistent field equations are solved by adaptive linear mixing. \\
{\em Restrictions:}  Spherical symmetry \\
{\em Unusual features:} \\
Radial integrators work for general potentials and meshes. No restriction to Coulombic or  self-consistent potentials; no restriction to uniform or exponential meshes. Outward Poisson integration. Fallback to bisection for robustness. \\
{\em Additional comments:}\\
{\em Running time:} For uranium, non-relativistic density functional
calculation execution time is around 0.6s for $10^{-6}\rm\,a.u.$ accuracy
in total energy on Intel Core i7 1.46GHz processor. \\
\end{small}

\section{Introduction}

Atomic structure calculations are a lynchpin of modern materials theory. 
They provide the basis for understanding the properties of individual atoms and 
play a central role in electronic structure calculations of larger, multi-atom systems 
such as molecules, nanostructures, and solids \cite{martin}. In the latter context, being a core computational kernel for larger-scale calculations, both accuracy and efficiency are of prime importance. Such atomic calculations, solving radial Schr\"odinger or Dirac equations, can be found at the heart of plasma physics calculations \cite{WilSS06}, all-electron electronic structure methods such as Korringa--Kohn--Rostoker (KKR) \cite{ButDG92}, linearized muffin tin orbital (LMTO) \cite{Skr84}, and linearized augmented planewave (LAPW) \cite{SinN06}, pseudo-atomic-orbital based methods \cite{ArtAD08}, \tit{ab initio} pseudopotential construction \cite{martin}, and projector augmented wave (PAW) \cite{Blo94}, relaxed-core PAW \cite{MarK06}, and all-electron pseudopotential (AEPP) \cite{vackarAEPP2} methods, among others.

Due to their central importance in the full range of electronic structure
calculations, from isolated atoms to solids, a number of atomic structure codes
have been developed over the decades, see, e.g.,
\cite{desclaux,hamann,grasp2k,FroeseFischer1991,atompaw,Jia2013,elk}. However,
most have been developed within, and remain closely tied to, the specific
larger-scale method of which they are a part, and hence are difficult to
separate and use for other purposes. Indeed, it was precisely our own need for
a robust and general solver for use in a finite-element~\cite{PasS05b} based
AEPP~\cite{vackarAEPP2} method that motivated the present work. Beyond
difficulty to extract, however, existing codes have widely differing
capabilities, efficiencies, robustness, and availability.
Table~\ref{table:solvers} lists features of several established codes. Some
solve the equations of density functional theory (DFT) while others solve
Hartree-Fock (HF). Some are relativistic, others not. Some can be converged to
high precision while others can be difficult to converge beyond the accuracies
required by the larger codes of which they are a part. Some allow any mesh
while others implement only exponential meshes. Some allow any potential while
others implement only singular all-electron potentials. Some implement
perturbation corrections for speed, others only bisection. Some are open
source, others not. Note that the table shows only shooting-type solvers
\cite{press07}, as we discuss in the present work. Other non-shooting type
approaches, see, e.g.,
\cite{Cayford1974,Tobin1975,Biegler-Konig1986,Andrae1997,Andrae2000,Andrae2001,FisZ09,Gra09}
and references therein, have been developed and implemented as well but have not yet found wide adoption in larger-scale electronic structure calculations due in part to robustness issues (e.g., spurious states) which can arise \cite{Gra09}. Note also that the table is intended only as a general guide: in the context of any given feature, ``No'' should be understood to mean only that we did not find it straightforward to implement in the referenced code.

\begin{table}
\begin{center}
{\scriptsize
\begin{tabular}{l c c c c c c c c}
Code                    & dftatom & Desclaux & atompp & grasp2k & MCHF & atompaw & PEtot & Elk \\
\hline
Reference               & This work & \cite{desclaux} & \cite{hamann} & \cite{grasp2k} & \cite{FroeseFischer1991} &
                                              \cite{atompaw} & \cite{Jia2013} & \cite{elk} \\
Open Source             & MIT & No  & No  & No  & No  & GPL & BSD & GPL \\
Schr\"odinger           & Yes & No  & Yes & No  & Yes & Yes & Yes & No  \\
Dirac                   & Yes & Yes & No  & Yes & No  & No  & Yes & Yes \\
DFT                     & Yes & No  & Yes & No  & No  & Yes & Yes & Yes \\
Any potential           & Yes & No  & Yes & No  & No  & No  & No  & Yes \\
Any mesh                & Yes & No  & No  & No  & No  & No  & No  & Yes \\
Outward Poisson         & Yes & No  & No  & No  & No  & No  & No  & No  \\
Perturb. correct.       & Yes & Yes & Yes & No  & Yes & Yes & Yes & No  \\
Bisection fallback      & Yes & No  & No  & No  & No  & No  & No  & No  \\
\hline
\end{tabular}
}
\caption{Features of various atomic structure solvers.}
\label{table:solvers}
\end{center}
\end{table}

In the present work, we bring together ideas from a host of atomic structure codes developed over many decades, and add new ideas to increase accuracy and robustness. We try to do this in such a way as to provide as general, flexible, and efficient a tool as possible, facilitating subsequent use in a range of applications areas. 
We develop and implement solvers for the radial Schr\"odinger, Dirac, and Kohn--Sham equations. The formulation allows general potentials and meshes: singular Coulomb-, finite pseudo-, or other potentials; uniform, exponential, or other meshes, as defined by nodal distributions and derivatives. For a given mesh type, convergence can be controlled systematically
by increasing the number of grid points. Radial integrations are carried out
using a combination of asymptotic forms, Runge-Kutta, and implicit Adams
methods. Eigenfunctions are determined by a combination of bisection and
perturbation methods for robustness and speed. Unlike other available codes
\cite{desclaux,hamann,grasp2k,FroeseFischer1991,atompaw,Jia2013,elk},
we employ an outward Poisson
integration to increase accuracy in the core region, allowing
absolute accuracies of $10^{-8}$ Hartree to be attained for total energies of
heavy atoms such as uranium. We show detailed convergence studies and
provide computational parameters to achieve accuracies commonly required
in practice. We present comparisons to analytic and current-benchmark density-functional
results \cite{nist,nistweb} for atomic number $Z$ = 1--92, verifying and refining current benchmarks. 

We provide an efficient, modular Fortran 95
implementation, \ttt{dftatom}, as open source, including
examples, tests, and wrappers for interface to other languages. 
The code consists of several
clearly arranged Fortran 95 modules, designed to be easy to apply for other purposes. 
The integration routines are independent of mesh, which can be specified as desired. Routines for common meshes are provided. Extensive tests are included to verify the correctness of results and serve as examples for a number of cases:
Coulombic, self consistent DFT, and non-singular potentials; exponential and hyperbolic meshes; atomic numbers $Z$ = 1--92; user-specified occupation numbers; and others. 
The code is opensource, available under the terms of the MIT license, 
from~\cite{dftatom}. C and Python wrappers are provided so that the code can be
easily used from other languages as well as interactively from Python.

The remainder of the paper is organized as follows. 
Section~\ref{equations} describes
the radial Schr\"odinger, Dirac, and Poisson equations, associated asymptotics,
and DFT equations to be solved. 
Section~\ref{methods} describes the numerical methods used to solve the radial
equations and eigenproblem, and meshes employed.
Section~\ref{results} shows Schr\"odinger and Dirac results for 
singular and nonsingular potentials, 
relativistic and nonrelativistic DFT calculations of uranium, convergence studies determining parameters for specified accuracies, and comparison to established benchmarks \cite{nist}. We conclude in Section~\ref{conclusions}.

\section{Equations}
\label{equations}

\subsection{Radial Schr\"odinger equation}
\label{schroedinger}

The 3D one-electron Schr\"odinger equation is given by
\begin{equation}
\left(-\half\nabla^2+V({\bf x})\right)\psi({\bf x}) = E\psi({\bf x}),
\end{equation}
in Hartree atomic units, as we use throughout. 
For a spherically symmetric potential
\begin{equation}
V({\bf x})=V(r),
\end{equation}
eigenstates of energy and angular momentum can be written in the form
\begin{equation}
\psi_{nlm}({\bf x})=R_{nl}(r)\,Y_{lm}\left({\bf x}\over r\right),
\label{psi}
\end{equation}
where $n$ is the principal quantum number, $l$ is the orbital angular momentum quantum number, and $m$ is the magnetic quantum number; whereupon it follows that $R_{nl}(r)$ satisfies the radial Schr\"odinger equation
\begin{equation}
\left(-\half r^2 R_{nl}'(r)\right)' + \left(r^2 V + \half l(l+1)\right)R_{nl}(r)
    =E r^2 R_{nl}(r).
\label{radial}
\end{equation}
The functions
$\psi_{nlm}({\bf x})$ and $R_{nl}(r)$ are normalized as
\begin{align}
\int |\psi_{nlm}({\bf x})|^2 \d^3 x &= 1, \\
\int_0^\infty R_{nl}^2(r) r^2 \d r &= 1.
\end{align}
The substitution
$P_{nl}(r) = rR_{nl}(r)$ and $Q_{nl}(r) = P_{nl}'(r) = R_{nl}(r) + rR_{nl}'(r)$
can be used to write the second-order radial
Schr\"odinger equation as a coupled set of first-order equations:
\begin{align}
\label{schroed_eq}
P_{nl}'(r) &= Q_{nl}(r), \\
Q_{nl}'(r) &= 2\left(V(r)-E + {l(l+1)\over 2r^2}\right)P_{nl}(r).
\end{align}
Such a first-order formulation facilitates solving both nonrelativistic Schr\"odinger and relativistic Dirac equations using the same techniques.
The normalization of $P(r)$ is
\begin{equation}
\int_0^\infty P_{nl}^2(r) \d r = 1.
\end{equation}
For a potential $V(r)$ which behaves like $-Z/r + V_0$ near the origin, 
as in the vicinity of the nucleus, the asymptotic behaviors of $P_{nl}$ and $Q_{nl}$ are~\cite{johnson}
\begin{align}
\label{sch_asympt_outward}
P_{nl}(r) &= r^{l+1},\\
\label{sch_asympt_outward2}
Q_{nl}(r) &= (l+1)r^l.
\end{align}
For large $r$, assuming $V(r) \to 0$ as $r\to\infty$, the asymptotic
is~\cite{johnson}:
\begin{align}
P_{nl}(r) &= e^{-\lambda r},\\
Q_{nl}(r) &= -\lambda P_{nl}(r),
\end{align}
where
\begin{equation}
    \lambda = \sqrt{-2E_{nl}}\,.
\end{equation}
As solutions of a homogeneous system, $P_{nl}$ and $Q_{nl}$ are unique only up to an arbitrary multiplicative constant. These small- and large-$r$ asymptotics provide starting values and derivatives for inward and outward numerical integrations.

\subsection{Radial Dirac equation}
\label{dirac}

The one-electron radial Dirac equation can be written as
\begin{align}
\label{dirac_eq1}
P_{n\kappa}'(r) &= -{\kappa\over r}P_{n\kappa}(r)+\left({E-V(r)\over c}+2c\right)Q_{n\kappa}(r), \\
\label{dirac_eq2}
Q_{n\kappa}'(r) &= -\left({E-V(r)\over c}\right)P_{n\kappa}(r)+{\kappa\over r}Q_{n\kappa}(r),
\end{align}
where $P_{n\kappa}(r)$ and $Q_{n\kappa}(r)$ are related to the usual large $g_{n\kappa}(r)$ and small $f_{n\kappa}(r)$ components
of the Dirac equation by
\begin{align}
P_{n\kappa}(r)&=rg_{n\kappa}(r), \\
Q_{n\kappa}(r)&=rf_{n\kappa}(r).
\end{align}
See for example~\cite{strange,zabloudil} for a discussion and derivation.
The solutions of the Dirac equation are labeled by quantum numbers $n$ and
$\kappa$,
where $n$ is the principal quantum number
and
$\kappa \ne 0$ is an integer.
Alternatively, they may be labeled by the triplet $n$, $l$, $s$,
where $l$ is the orbital angular momentum
and $s=\pm1$ distinguishes the $j=l\pm\half$ states ($j$ is the total angular
momentum). $\kappa$ is then determined by
\begin{equation}
\kappa    = \begin{cases}
                -l-1 & \text{for $j=l+\half$, i.e. $s=+1$}, \\
                \quad l   & \text{for $j=l-\half$, i.e. $s=-1$}.
            \end{cases}
\end{equation}
The energy $E$ does not contain the electron rest mass energy, so it can be 
compared directly to the corresponding nonrelativistic energy of the Schr\"odinger
equation. 
The normalization of $P_{n\kappa}(r)$ and $Q_{n\kappa}(r)$ is
\begin{equation}
\int_0^\infty (P_{n\kappa}^2(r) + Q_{n\kappa}^2(r)) \d r = 1.
\end{equation}

For potentials of the form $V(r) = -Z/r + Z_1 + O(r)$,
the asymptotic behavior at the origin for $Z \ne 0$ (Coulombic) 
is~\cite{Grant2008,zabloudil}
\begin{align}
P_{n\kappa}(r) &= r^{\beta}, \\
Q_{n\kappa}(r) &= r^{\beta} {c(\beta+\kappa)\over Z},
\end{align}
where
\begin{equation}
\beta=\sqrt{\kappa^2-\left(Z\over c\right)^2}.
\end{equation}
For $Z=0$ (nonsingular) the asymptotic is~\cite{Grant2008}, for $\kappa < 0$
\begin{align}
\label{dirac_asympt_outward_nonsing1}
P_{n\kappa}(r) &=  r^{l+1}, \\
\label{dirac_asympt_outward_nonsing1b}
Q_{n\kappa}(r) &=  r^{l+2} {E + Z_1 \over c(2l+3)},
\end{align}
and for $\kappa > 0$
\begin{align}
\label{dirac_asympt_outward_nonsing2}
P_{n\kappa}(r) &= -r^{l+2} {E + Z_1 \over c(2l+1)}, \\
\label{dirac_asympt_outward_nonsing2b}
Q_{n\kappa}(r) &=  r^{l+1}.
\end{align}
For large $r$, assuming $V(r) \to 0$ as $r\to\infty$, the asymptotic
is~\cite{johnson}
\begin{align}
    P_{n\kappa}(r) &= e^{-\lambda r},\\
    Q_{n\kappa}(r) &= - \sqrt{-{E\over E + 2c^2}} P_{n\kappa}(r),\\
\end{align}
where
\begin{equation}
\lambda = \sqrt{c^2 -{(E+c^2)^2\over c^2}} =\sqrt{-2E - {E^2\over c^2}}.\\
\end{equation}
As in the case of the Schr\"odinger equation, $P_{n\kappa}$ and $Q_{n\kappa}$
are unique only up to an arbitrary multiplicative constant, and the above
asymptotics provide starting values and derivatives for inward and outward
numerical integrations.

\subsection{Poisson equation}
\label{sec:poisson}

The 3D Poisson equation for the Hartree potential $V_H$ 
due to electronic density $n$ is given by
\begin{equation}
\nabla^2V_H({\bf x}) = -4\pi n({\bf x}).
\end{equation}
For a spherical density $n({\bf x}) = n(r)$, this becomes
\begin{equation}
{1\over r^2}(r^2 V_H')' = V_H''(r) + {2\over r}V_H'(r) = -4\pi n(r),
\label{poisson:eq}
\end{equation}
where $n(r)$ is the radial particle (number) density,
normalized such that
\begin{equation}
N = \int n({\bf x}) \d^3 x = \int_0^\infty 4\pi n(r) r^2 \d r\,,
\label{poisson:norm}
\end{equation}
where $N$ is the electron number.
Substituting (\ref{poisson:eq}) into (\ref{poisson:norm}) and integrating, we obtain
\begin{equation}
\lim_{r\to\infty} r^2 V_H'(r) = -N,
\end{equation}
from which it follows that the asymptotic behavior of $V_H'(r)$ is
\begin{equation}
V_H'(r) = - {N\over r^2}, \qquad r \to \infty.
\label{poisson:V_H'}
\end{equation}
Integrating (\ref{poisson:V_H'}) and requiring $V_H \to 0$ as $r \to \infty$ then gives the corresponding asymptotic behavior for $V_H(r)$:
\begin{equation}
V_H(r) = {N\over r}, \qquad r \to \infty.
\end{equation}
For small $r$, the asymptotic behavior can be obtained by expanding $n(r)$ about $r=0$: $n(r) = \sum_{j=0}^\infty {c_j r^j}$. Substituting into Poisson equation \eqref{poisson:eq} gives
\begin{equation}
(r^2 V_H')' = -4\pi \sum_{j=0}^\infty {c_j r^{j+2}}.
\end{equation}
Integrating and requiring $V_H(0)$ finite then gives
\begin{equation}
\label{eq:dVsmallr}
V_H'(r) = -4\pi \sum_{j=0}^\infty { c_j \frac{r^{j+1}}{j+3} },
\end{equation}
with linear leading term; so that we have
\begin{equation}
\label{eq:dVsmallr2}
V_H'(r) \propto r, \qquad r \to 0.
\end{equation}
Integrating \eqref{eq:dVsmallr} then gives
\begin{equation}
\label{eq:Vsmallr}
V_H(r) = -4\pi \sum_{j=0}^\infty { c_j \frac{r^{j+2}}{(j+2)(j+3)} } + C,
\end{equation}
with leading constant term $C = V_H(0)$ determined by Coulomb's law:
\begin{equation}
\label{VH0}
V_H(0) = 4 \pi \int_0^\infty {r n(r) \d r}.
\end{equation}
Finally, from \eqref{eq:dVsmallr2} we have that
\begin{equation}
\label{dVH0}
V_H'(0) = 0.
\end{equation}
The above asymptotics provide starting values and derivatives for inward and outward numerical integrations.


\subsection{Kohn--Sham equations}
\label{DFT}

The Kohn--Sham equations consist of the radial Schr\"odinger or Dirac equations above
with an effective potential $V(r) = V_\inrm(r)$ given by (see, e.g.,~\cite{martin})
\begin{equation}
V_\inrm = V_H + V_{xc} + v,
\end{equation}
where $V_H$ is the Hartree potential given by the solution of the radial
Poisson equation (\ref{poisson:eq}), $V_{xc}$ is the exchange-correlation
potential and $v = -{Z\over r}$ is the nuclear potential.

The total energy is given by
\begin{equation}
E[n] = T_s[n] + E_H[n] + E_{xc}[n] + V[n],
\end{equation}
the sum of kinetic energy
\begin{equation}
T_s[n] = \sum_{nl} f_{nl} \epsilon_{nl} -4\pi\int V_\inrm(r) n(r) r^2 \d r,
\end{equation}
where $\epsilon_{nl}$ are the Kohn-Sham eigenvalues,
Hartree energy
\begin{equation}
E_H[n] = 2\pi\int V_H(r) n(r) r^2 \d r,
\end{equation}
exchange-correlation energy
\begin{equation}
E_{xc}[n] = 4\pi\int \epsilon_{xc}(r; n) n(r) r^2 \d r,
\end{equation}
where $\epsilon_{xc}(r; n)$ is the exchange and correlation energy density,
and Coulomb energy
\begin{equation}
V[n] = 4\pi\int v(r) n(r) r^2 \d r
    = -4\pi Z\int n(r) r \d r;
\end{equation}
with electronic density in the nonrelativistic case given by
\begin{equation}
    \label{eq-schr-density}
    n(r) = {1\over4\pi} \sum_{nl} f_{nl} {P_{nl}^2(r)\over r^2},
\end{equation}
where $P_{nl}$ is the radial wavefunction (Eq.~\eqref{schroed_eq}) and
$f_{nl}$ the associated electronic occupation. In the relativistic case, the
electronic density is given by
\begin{equation}
    \label{eq-dirac-density}
    n(r) = {1\over4\pi} \sum_{nls} f_{nls}
    {P_{nls}^2(r) + Q_{nls}^2(r) \over r^2},
\end{equation}
where $P_{nls}$ and $Q_{nls}$ are the two components of the Dirac
solution (Eqs.~\eqref{dirac_eq1},~\eqref{dirac_eq2})
and $f_{nls}$ is the occupation. 
In both cases above, $n(r)$ is the electronic particle density [electrons/volume], everywhere positive, as distinct from the electronic charge density $\rho(r)$ [charge/volume]: $\rho(r) = -n(r)$ in atomic units.

These equations are solved self-consistently (see, e.g.,~\cite{martin}): an initial density $n_\inrm$ and corresponding potential $V_\inrm$ are constructed; the Schr\"odinger or Dirac equation is solved to determine wavefunctions $R_{nl}$ or spinor components $P$ and $Q$, respectively; from these a new density $n_\outrm$ and potential $V_\outrm$ are constructed; and from these, a new input density $n_\inrm$ and potential $V_\inrm$ are constructed. The process is continued until the difference of $V_\inrm$ and $V_\outrm$ and/or $n_\inrm$ and $n_\outrm$ is within a specified tolerance, at which point \tit{self-consistency} is achieved. This fixed point iteration is known as the \tit{self-consistent field} (SCF) iteration. We employ an adaptive linear mixing scheme, with optimized weights for each component of the potential to construct new input potentials for successive SCF iterations. In order to reduce the number of SCF iterations, we use a 
Thomas-Fermi (TF) approximation~\cite{oulne} for the initial density and potential:
\begin{equation}
V(r) = -{Z_\eff(r)\over r},
\end{equation}
where
\begin{gather}
\label{tf-solution}
Z_\eff(r) = Z (1 + \alpha \sqrt{x} + \beta x e^{-\gamma \sqrt{x}})^2
    e^{-2 \alpha \sqrt{x}}, \\
x = r \left(128 Z\over 9\pi^2\right)^{1\over3}, \\
\alpha = 0.7280642371, \\
\beta = -0.5430794693, \\
\gamma = 0.3612163121. 
\end{gather}
The corresponding charge density is then
\begin{equation}
\label{tf:rho}
\rho(r) = -{1 \over 3 \pi^2} \left(-2 V(r)\right)^{3\over2}.
\end{equation}
For simplicity, we consider only local-density approximation (LDA) and relativistic local-density approximation (RLDA) exchange and correlation potentials here. More sophisticated functionals can be readily incorporated. Our \ttt{dftatom} implementation uses the same parameterization as in NIST benchmarks~\cite{nist}: 
\begin{equation}
V_{xc}(r;n) = {\d \over \d n}\left(n\epsilon_{xc}^{LD}(n)\right),
\end{equation} 
where the exchange and correlation energy density $\epsilon_{xc}^{LD}$ 
can be written as~\cite{martin} 
\begin{equation}
\epsilon_{xc}^{LD}(n)=\epsilon_x^{LD}(n)+\epsilon_c^{LD}(n),
\end{equation} 
with electron gas exchange term~\cite{martin} 
\begin{equation}
\epsilon_x^{LD}(n)=-{3\over4\pi}(3\pi^2 n)^{1\over3}
\end{equation} 
and Vosko-Wilk-Nusair (VWN)~\cite{VWN} correlation term
\begin{multline}
\epsilon_c^{LD}(n)\approx {A\over2}\left\{
\log\left(y^2\over Y(y)\right)
+{2b\over Q}\arctan\left(Q\over 2y+b\right) \right. \\ \left.
-{by_0\over Y(y_0)}\left[
   \log\left((y-y_0)^2\over Y(y)\right)
   +{2(b+2y_0)\over Q}\arctan\left(Q\over 2y+b\right)
   \right]
\right\},
\end{multline} 
in which $y=\sqrt{r_s}$, $Y(y)=y^2+by+c$, $Q=\sqrt{4c-b^2}$, $y_0=-0.10498$, 
$b=3.72744$, $c=12.9352$, $A=0.0621814$, and 
\begin{equation}
r_s=\left(3\over4\pi n\right)^{1\over3}
\end{equation} 
is the Wigner-Seitz radius, which gives the mean distance between electrons. 
In the relativistic (RLDA) case, a correction to the LDA 
exchange energy density and potential is given 
by MacDonald and Vosko~\cite{donald:RLDA}: 
\begin{gather}
\epsilon_x^{RLD}(n) = \epsilon_x^{LD}(n) R, \\
R = 1-{3\over2}\left(\beta\mu-\log(\beta+\mu)\over\beta^2\right)^2, \notag \\
V_{x}^{RLD}(n) = V_{x}^{LD}(n) S, \\
S = {3\log(\beta+\mu)\over 2 \beta \mu} - \half, \notag
\end{gather} 
where $\mu=\sqrt{1+\beta^2}$ and $\beta={(3\pi^2n)^{1\over3}\over c}
=-{4\pi \epsilon_x^{LD}(n)\over 3c}$. 

\section{Methods of solution}
\label{methods}

In this section, we discuss the radial integration methods, meshes, and eigenfunction isolation methods employed in the Kohn--Sham solution.

\subsection{Runge-Kutta and Adams methods}

To allow general, nonuniform meshes, all methods first transform equations on a general mesh $R(t)$, $1 \le t \le N + 1$, to equations on a uniform mesh $t$ with step size $h=1$.
If the solution on a general mesh is $P(r)$ and the transformed solution on the
uniform mesh is $u(t)$ then
\begin{align}
   u(t) &= P(R(t)) \\
   u'(t) &= {\d u\over\d t} = {\d P\over\d R} R'(t) \\
\end{align}
The methods below require the
values $u(t)$ and derivatives $u'(t)$ on the uniform mesh, which we
express in terms of the values $P(r)$ and derivatives $P'(r)$ on the
general mesh, and derivative $R'(t)$ of the function defining the general mesh.

The Runge-Kutta family of methods 
for the numerical integration of ordinary differential equations 
require the values of dependent and independent variables 
in the interior of the element (step). 
For example, the 4th-order Runge-Kutta
(RK4) step for an equation of the form $y' = f(x, y)$ is:
\begin{align}
y_{i+1} &= y_i + {1\over 6} (k_1 + 2k_2 + 2k_3 + k_4), \\
k_1     &= f(x_i, y_i), \\
k_2     &= f(x_{i + \half}, y_i + \half k_1), \\
k_3     &= f(x_{i + \half}, y_i + \half k_2), \\
k_4     &= f(x_{i+1}, y_i + k_3).
\end{align}
Implicit Adams methods, on the other hand, use an extrapolation formula to
advance the solution to the next grid point and then an interpolation formula
to correct the value (predictor-corrector) using the grid points only. 
The 4th-order Adams outward extrapolation is given by
\begin{equation}
y_{i+1} = y_i + {1\over 24} (55y_i' - 59y_{i-1}' + 37y_{i-2}' - 9y_{i-3}')
\end{equation}
and interpolation by
\begin{equation}
y_{i+1} = y_i + {1\over 24} (9y_{i+1}' + 19y_i' -5 y_{i-1}' + y_{i-2}').
\end{equation} 
Since the Schr\"odinger and Dirac equations are linear 
(for given effective potential in each SCF iteration), 
the interpolation can be determined analytically, thus eliminating 
predictor-corrector iterations, speeding up the calculation considerably.
Both Schr\"odinger and Dirac equations can be written as
\begin{equation}
    \begin{pmatrix}
        P' \\
        Q' \\
    \end{pmatrix}
    =
    C
    \begin{pmatrix}
        P \\
        Q \\
    \end{pmatrix},
\end{equation}
where for the Schr\"odinger equation, the matrix $C$ is given by
\begin{equation}
    C(i) = \begin{pmatrix}
        0 & 1 \\
        {l(l+1)\over R^2(i)} + 2(V(i)-E) & 0 \\
        \end{pmatrix}
\end{equation}
and for the Dirac equation,
\begin{equation}
    C(i) = \begin{pmatrix}
        -{\kappa\over R(i)} & {E-V(i)\over c} + 2c \\
        -{E-V(i)\over c} & {\kappa\over R(i)} \\
        \end{pmatrix}.
\end{equation}
Then for outward integration, analytic interpolation gives:
\begin{align}
    \lambda & = {9\over 24}, \\
    \Delta  & = 1 + \lambda^2 R'^2(i+1) \det C(i+1), \\
    M       & = {1\over \Delta}\left(\begin{pmatrix}
        1 & 0 \\
        0 & 1 \\
    \end{pmatrix}  + \lambda R'(i+1)
    \begin{pmatrix}
        -C_{22}(i+1) & C_{12}(i+1) \\
        C_{21}(i+1) & -C_{11}(i+1) \\
        \end{pmatrix}
    \right), \\
    \begin{pmatrix}
        u_1(i+1) \\
        u_2(i+1) \\
    \end{pmatrix} & = M
        \begin{pmatrix}
            u_1(i) + {1\over 24}(19 u_1'(i) - 5 u_1'(i-1) + u_1'(i-2)) \\
            u_2(i) + {1\over 24}(19 u_2'(i) - 5 u_2'(i-1) + u_2'(i-2)) \\
        \end{pmatrix}.
\end{align}
See~\cite{johnson} for a detailed discussion.

For inward integration, the 4th order Adams inward extrapolation is given by
\begin{equation}
y_{i-1} = y_i - {1\over 24} (55y_i' - 59y_{i+1}' + 37y_{i+2}' - 9y_{i+3}')
\end{equation}
and interpolation by
\begin{equation}
y_{i-1} = y_i - {1\over 24} (9y_{i-1}' + 19y_{i}' - 5y_{i+1}' + y_{i+2}'),
\end{equation}
in which case, analytic interpolation gives:
\begin{align}
    \lambda & = -{9\over 24}, \\
    \Delta  & = 1 + \lambda^2 R'^2(i-1) \det C(i-1), \\
    M       & = {1\over \Delta}\left(\begin{pmatrix}
        1 & 0 \\
        0 & 1 \\
    \end{pmatrix}  + \lambda R'(i-1)
    \begin{pmatrix}
        -C_{22}(i-1) & C_{12}(i-1) \\
        C_{21}(i-1) & -C_{11}(i-1) \\
        \end{pmatrix}
    \right), \\
    \begin{pmatrix}
        u_1(i-1) \\
        u_2(i-1) \\
    \end{pmatrix} & = M
        \begin{pmatrix}
            u_1(i) - {1\over 24}(19 u_1'(i) - 5 u_1'(i+1) + u_1'(i+2)) \\
            u_2(i) - {1\over 24}(19 u_2'(i) - 5 u_2'(i+1) + u_2'(i+2)) \\
        \end{pmatrix}.
\end{align}

In practice, we use RK4 to compute the first 4 points needed by implicit Adams, then switch to the more accurate implicit Adams method to compute the remaining points.

\subsection{Mesh}
\label{mesh}

The methods discussed here can use any desired mesh: uniform, hyperbolic, exponential, or other. In practice, however, as we show below, most atomic structure codes use some form of exponential (or ``logarithmic'') mesh, as this gives an efficient concentration of mesh points in the vicinity of the Coulomb singularity, where wavefunctions, densities, and potentials vary most rapidly. Given their particular prevalence, we discuss exponential meshes below, and provide associated routines in the \ttt{dftatom} distribution.

Every exponential mesh can be written as a function of exactly four parameters,
$r_\min$, $r_\max$, $a$, and $N$, as follows:
\begin{align}
\label{expmesh}
R_i &= \alpha \left(e^{\beta(i-1)}-1\right) + r_\min, \\
\alpha &= {r_\max - r_\min \over e^{\beta N} - 1}, \\
\beta &= {\log a \over N-1}.
\end{align}
for $i=1, 2, \dots, N+1$. 
Note that in the above we have explicitly included an arbitrary shift of origin. Without this, three parameters are of course sufficient to specify an exponential mesh. As we elaborate below, however, the flexibility to choose the initial point independent of other mesh parameters affords additional efficiency when employing asymptotic expressions for initial values.
In the above form, all parameters have direct physical meaning: the
mesh starts at $R_1 = r_\min$ and ends at $R_{N+1} = r_\max$; as shown below,
the parameter
$a$ is the ratio of rightmost to leftmost element (mesh interval) lengths (determining the mesh gradation), and $N$ is the number of elements in the
mesh. The ratio $q$ of any two successive elements can be calculated from
\eqref{expmesh}:
\begin{equation}
q = {R_{n+1} - R_n \over R_n - R_{n-1}} = e^\beta = a^{1\over N-1}.
\end{equation}
It follows that the mesh gradation $a$ can be expressed
using the ratio $q$ as $a = q^{N-1}$ and thus $a$ is the ratio of 
rightmost to leftmost element lengths.

The main advantage of this parametrization is that $r_\min$ and $r_\max$ 
are decoupled from the mesh gradation and number of elements and can thus 
 be varied independently. 
Once $r_\min$ is determined to provide sufficiently accurate 
asymptotic values, there is still freedom to change the gradation of the mesh by 
changing $a$. Then, after determining the optimal $a$, mesh convergence 
can be achieved by simply increasing $N$. By having the option to optimize $a$ independent of both $r_\min$ and $N$, one can 
reduce the number of elements $N$ required for a given accuracy, thus speeding
up the calculation.

Exponential meshes in common use are special cases of the above more general one. 
For example the first and second meshes in~\cite{nist} are given by
\begin{equation}
\label{expmesh1}
r_i = r_\min \left(r_\max\over r_\min\right)^{(i-1)/N}.
\end{equation}
Here, the meaning of $i$, $N$,
$r_\min$, and $r_\max$ is the same as in \eqref{expmesh}. This form is
a special case of \eqref{expmesh} with
$a = \left(r_\max/r_\min\right)^{N-1\over N}$. As
such, the mesh is determined by three parameters $r_\min$,
$r_\max$, and $N$, and the mesh gradation $a$ depends on $N$ and the fraction
$r_\max/r_\min$. Note that this mesh does not allow $r_\min=0$ (the
mesh gradation would become infinite). For uranium, the first mesh has
$N=15788$, $r_\min={1\over 160\times92}$, and $r_\max=50$;
from which it follows that $a=7.353705\times 10^5$. The second mesh has
$N=8000$, $r_\min={10^{-6}\over 92}$, and $r_\max={800\over \sqrt{92}}$;
from which it follows $a=7.651530\times 10^9$.

The third mesh in~\cite{nist} is given by
\begin{equation}
\label{expmesh3}
r_i = A(e^{B i} - 1).
\end{equation}
The meaning of $i$ is the same as in \eqref{expmesh}.
The mesh is determined by three parameters
$A$, $B$, and $N$. This is a special case of \eqref{expmesh}
with $r_\min=A(e^B - 1)$, 
$r_\max = A \left(e^{B (N+1)}-1\right)$, 
and $a=e^{B(N-1)}$. For uranium, the values in~\cite{nist} are
$A={4.34\over 92}\times 10^{-6}$, $B=0.002304$, and $N=9019$; from
which it follows that 
$r_\min=1.088140\times 10^{-10}$, $r_\max=50.031620$, and
$a=1.055702\times 10^9$.

The fourth mesh in~\cite{nist} is given by:
\begin{equation}
\label{expmesh4}
\rho_i = \log R_i,
\end{equation}
where $\rho_i$ is a uniform mesh with $N$ elements such that $R_1=r_\min$
and $R_{N+1} = r_\max$. It follows that
$\rho_i = {i-1\over N}\log{r_\max\over r_\min} + \log r_\min$.
Then substituting into \eqref{expmesh4} and simplifying, the expression
\eqref{expmesh1} is obtained. As such, the fourth mesh is equivalent to the first.
For uranium, the values in~\cite{nist} are $N=2837$, $r_\min={10^{-6}\over 92}$, and $r_\max=50$;
from which it follows that $a=4.564065\times 10^9$.

To summarize, the parameters of the four meshes for uranium in~\cite{nist} are then as follows.

\begin{verbatim}
  | r_min             r_max         a            N
------------------------------------------------------
1 | 6.79347826087e-05 50.0          7.353705e+05 15788
2 | 1.08695652174e-08 83.4057656228 7.651530e+09 8000
3 | 1.08814001246e-10 50.0316203306 1.055702e+09 9019
4 | 1.08695652174e-08 50.0          4.564065e+09 2837
\end{verbatim}

In Section~\ref{sec:U}, we determine optimal parameters for LDA and RLDA
calculations, as shown in Table~\ref{table:meshreference}. Here, the given 
$r_\min$, $r_\max$, and $a$ can be used for all atoms,
while the given $N$ is sufficient 
for $10^{-6}$ a.u.\ accuracy in total energy for uranium; it can be increased for higher
accuracy and decreased for lower atomic numbers $Z$ to improve speed
while retaining $10^{-6}$ accuracy, as we show in Section~\ref{sec:U}.

\begin{table}
\begin{center}
\begin{tabular}{c r r r r}
 & \multicolumn{1}{c}{$r_\min$} & \multicolumn{1}{c}{$r_\max$} &
        \multicolumn{1}{c}{$a$} & \multicolumn{1}{c}{$N$} \\
\hline
LDA  & 1.0e-07 &  50.0   & 2.7e+06 & 4866 \\
RLDA & 1.0e-08 &  50.0   & 6.2e+07 & 5268 \\
\end{tabular}
\caption{Optimal mesh parameters for $10^{-6}$ a.u.\ accuracy in total energy of uranium.}
\label{table:meshreference}
\end{center}
\vspace{0.1in}
\end{table}

\subsection{Shooting method for Schr\"odinger \& Dirac eigenproblems}
\label{eigenproblem}

The required eigenfunctions are determined by the shooting method~\cite{press07}: i.e., eigenvalues are guessed, radial integrations are performed, guesses are updated, and the process is repeated until convergence is achieved to the desired tolerance. For efficiency, we employ a combination of inward and outward integrations, bisection for robustness, and perturbation theory for efficiency once the solution is sufficiently close to convergence.

The outward integration starts at $r_\min$ using the asymptotic from 
Section~\ref{equations} 
with $r=r_\min$, then RK4 is used for the first four steps, then 
Adams for the rest. For the inward integration, a starting point $r_m$ is determined 
such that 
$e^{-\lambda r_m}$ is on the order of machine epsilon (e.g., $\sim 10^{-16}$ in double precision), 
with $\lambda$ determined from
the large-$r$ asymptotic for the Schr\"odinger or Dirac equation, as appropriate (Section~\ref{equations}). Then $e^{-\lambda r_m}$ is used for the last four
points up to $r_m$ and inward Adams is used for the rest.

The outward integration is performed up to the classical turning point $r_\ctp$,
defined by $V(r_\ctp) = E$. Bisection is used to converge the energy to the point that 
the associated radial solution has the correct number of nodes ($n-l-1$). After the
correct number of nodes is obtained, two approaches are considered to complete 
the eigenfunction computation.

The first approach is bisection:
the outward integration is continued from $r_\ctp$ to the rest of the domain and the total number
of nodes is counted. If the
number is strictly greater than $n-l-1$, then the energy $E$ is above the
eigenvalue, otherwise it is below the eigenvalue. Bisection is used to refine the 
eigenvalue to specified accuracy.

The second approach uses a perturbation correction for the energy, using both
inward and outward integrations. It typically converges an order of magnitude faster than outward-only bisection once the correct number of nodes is obtained. For the Schr\"odinger equation, the following correction is used~\cite{johnson,Ridley1955,Douglas1956,Hartree1957}:
\begin{equation}
E_2 \approx E_1
    + {P(r_\ctp)(Q(r_\ctp^-)-Q(r_\ctp^+))\over
        2\int^\infty_0 P^2(r) \,\d r},
\end{equation}
where $E_1$ is the energy used to calculate $P(r)$ and $Q(r)$. The inward
integration is multiplied by a constant so that it matches (in value) the
outward integration at the classical turning point $r_\ctp$. This leaves a
jump in the derivative of the outward integration $Q(r_\ctp^-)$ and inward
integration $Q(r_\ctp^+)$. $E_2$ is the new energy for the next iteration.
Usually around five iterations are sufficient to achieve $10^{-13}$ accuracy in
energy.

For Dirac equation, the following correction is used~\cite{johnson,Mayers1957}:
\begin{equation}
E_2 \approx E_1
    + c {P(r_\ctp)(Q(r_\ctp^-)-Q(r_\ctp^+))\over
        \int^\infty_0 \left( P^2(r) +Q^2(r) \right) \,\d r}.
\end{equation}

After the correct number of nodes is determined, the perturbation correction is tried first. If it fails to converge (as can occur, e.g., for a too-confined domain), we fall back to outward-only bisection for robustness. For positive energies, the perturbation correction is inapplicable and outward-only bisection is used exclusively. 
Physically, the outward-only bisection method solves the atom in
an infinite potential well of the size of the domain while the inward-outward perturbation method
solves an atom in an infinite domain (by virtue of the asymptotic inward
integration). As such, for large $r_\max$, the two methods converge to the same
value, but for small $r_\max$ the eigenvalues can differ between methods.
(An $r_\max$ study is needed to determine sufficient $r_\max$, as we show in Section~\ref{results}).

All normalization and other integrals are calculated on $[r_\min, r_\max]$,
avoiding the Coulomb singularity at $r=0$. Hence, convergence with respect to
$r_\min$ (typically $\sim 10^{-8}\rm\,a.u.$) must be checked, as we show in
Section~\ref{results}. 

\subsection{Outward Poisson integration}

Since asymptotics are known at both ends of the domain (Section~\ref{sec:poisson}), 
it is possible to integrate the Poisson equation either inward or outward. 
 Unlike other available codes
\cite{desclaux,hamann,grasp2k,FroeseFischer1991,atompaw,Jia2013,elk},
we employ outward integration here in order to more accurately resolve rapid 
variations in core region. This better resolves the most tightly bound states (e.g., $1s$ and $2s$), in particular.

Since charge is omitted around the Coulomb singularity at $r=0$ ($0< r < r_\min$), 
we take as initial conditions, consistent with Eqs.~\eqref{VH0} and \eqref{dVH0}: 
\begin{align}
V_H(r_\min) &= 4\pi\int_{r_\min}^{r_\max} r n(r) \d r, \\
V_H'(r_\min) &= 0.
\end{align}
Then RK4 is used for the first four points, after which the Adams method
is used with predictor-corrector for the rest. Here again, convergence with respect to $r_\min$ must be checked, as we show in Section~\ref{results}.

Having a precise Poisson solver is crucial to resolve all Kohn--Sham states
accurately. With this outward solver, Kohn--Sham eigenvalues are typically an 
order of magnitude more precise than the total energy, with the most pronounced effect on the more tightly bound states. This is in contrast to other solvers~\cite{nist} which typically resolve eigenvalues less accurately than total energies.

Outward integration of the Poisson equation has been considered previously \cite{Torrance1934,Yost1940}. However, rather than solving the differential equation, these authors compute $V_H(r)$ using
the integral formulation, which has implicit the initial conditions 
$V_H(r_\min) = V_H'(r_\min) = 0$, shifting energies by
$4\pi\int_{r_\min}^{r_\max} r n(r) \d r$, subtracted subsequently. 
Outward integration of the differential equation has been discussed in \cite{Hartree1957}, where integration of the differential equation for
$Y(r) = r V_H(r)$ is considered with initial conditions $Y(r) = Y'(r) = 0$, 
with subtraction of the associated energy shift subsequently.

\section{Results}
\label{results}

\subsection{Analytic test case: $V = -Z/r$}

To verify the accuracy of the solver and establish convergence indicators, 
the Schr\"odinger and Dirac equations are solved for potential $V = -Z/r$ with $Z=92$, 
for which analytic results are available.
$r_\min$, $r_\max$, and $N$ convergence studies are run for all
eigenvalues with $n \le 7$. In the resulting plots, $E-E_\mrm{prev}$ gives 
the change in energy with change of parameter being studied: e.g., as 
$r_\min$ is decreased toward convergence, $r_\min = 10^{-4}, 10^{-5}, \ldots, 10^{-14}$, 
$E-E_\mrm{prev}$ at $r_\min = 10^{-7}$ is the difference of computed energies at $r_\min = 10^{-7}$ and $r_\min = 10^{-6}$. 
$E-E_\mrm{conv}$ is the difference of the computed energy from the fully converged value: e.g., $E-E_\mrm{conv}$ at $r_\min=10^{-7}$ is the difference of energy at $r_\min=10^{-7}$ from the fully converged value at $r_\min$ sufficiently small that reducing it further does not reduce the error further (due to finite precision).
Note that convergence plots omit points at converged values
of parameters ($r_\min$, $r_\max$, $N$) because energy differences
$E-E_\mrm{conv}$ are identically zero there, and so cannot be plotted on the
log scale.

\begin{figure}
\centering
\subfloat[$r_\min$ study]{
\label{fig:Z_rmin}
\includegraphics[width=0.5\linewidth]{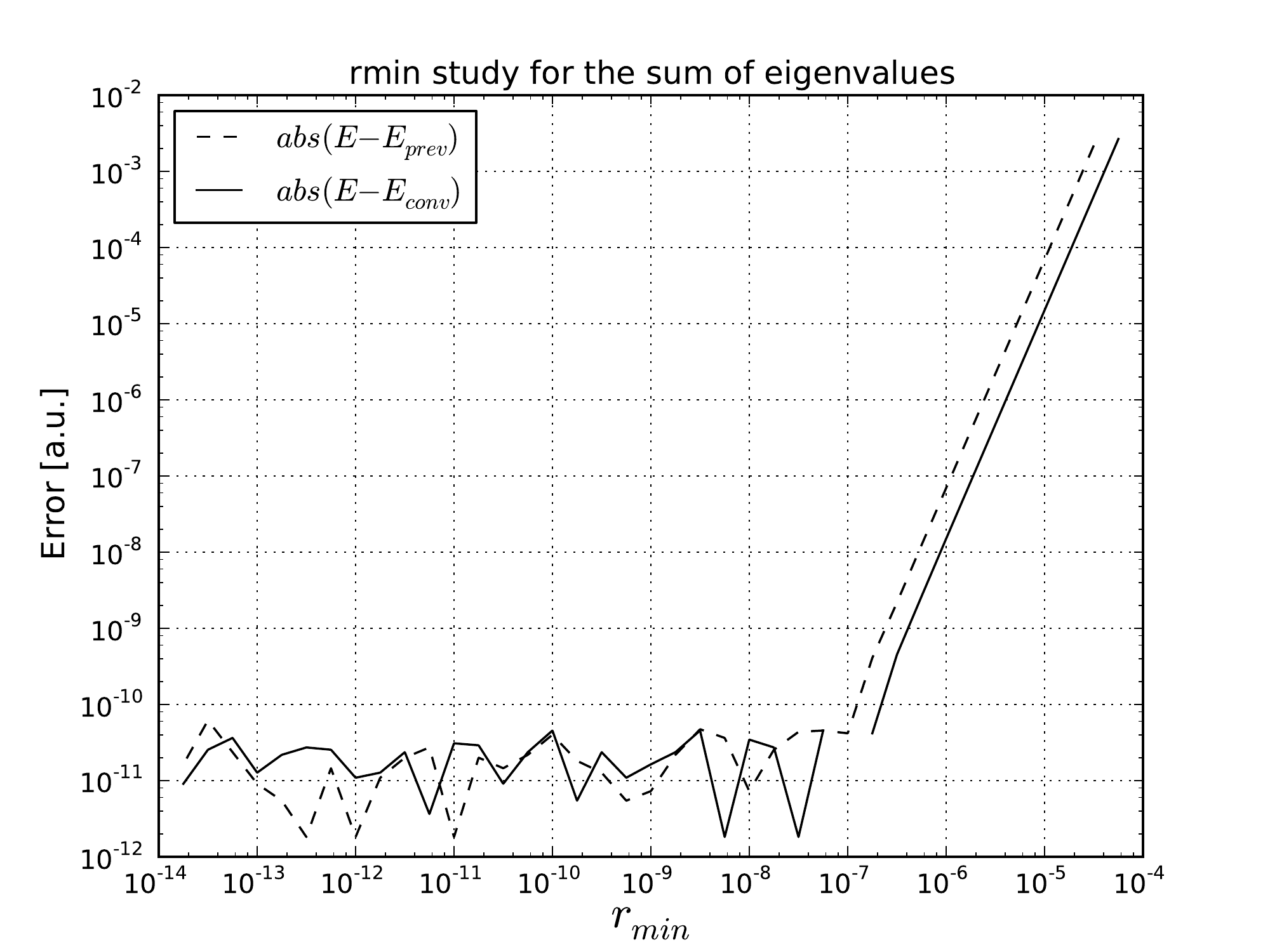}
}

\subfloat[$r_\max$ study]{
\label{fig:Z_rmax}
\includegraphics[width=0.5\linewidth]{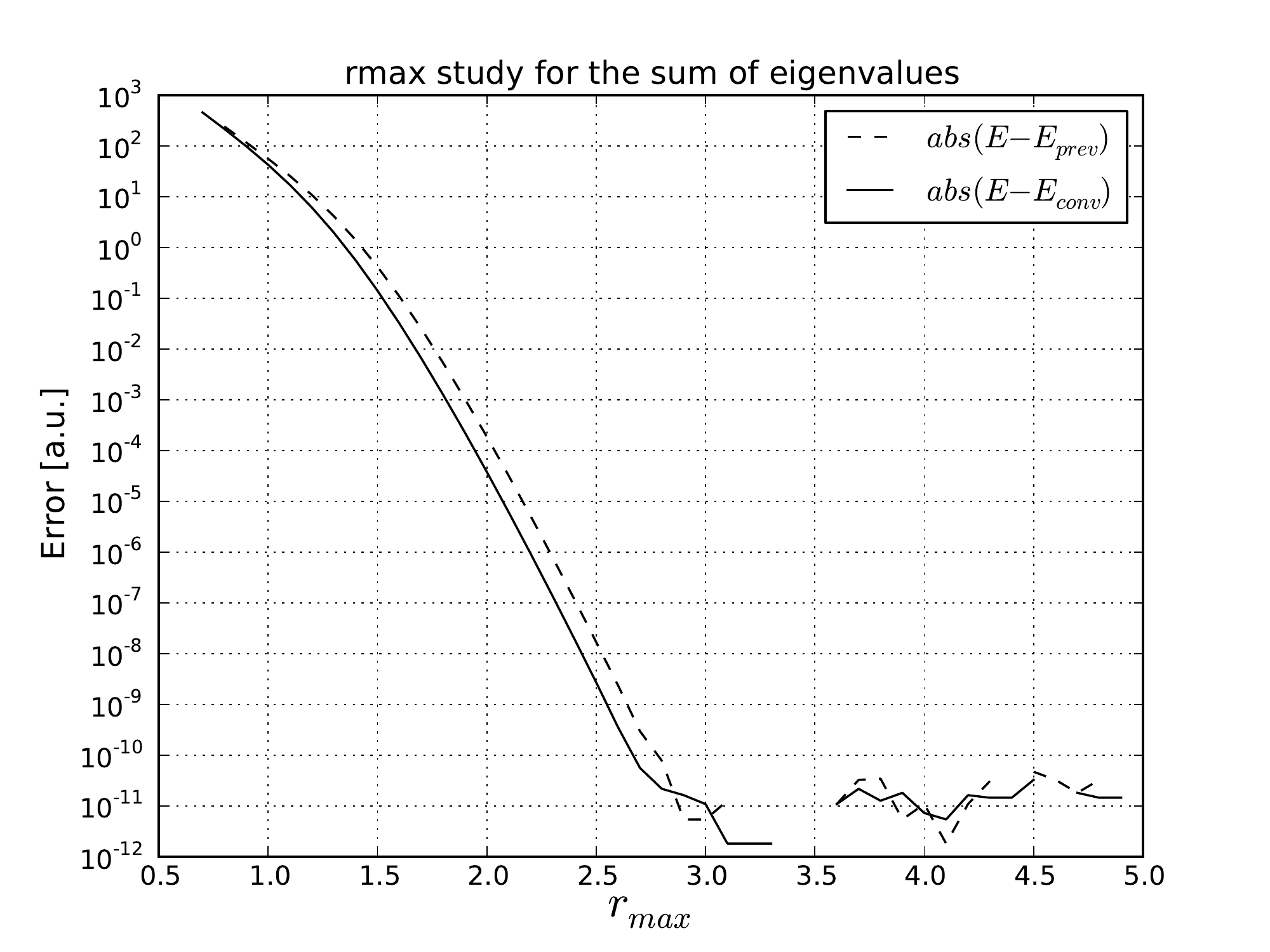}
}

\subfloat[$N$ study]{
\label{fig:Z_n_study}
\includegraphics[width=0.5\linewidth]{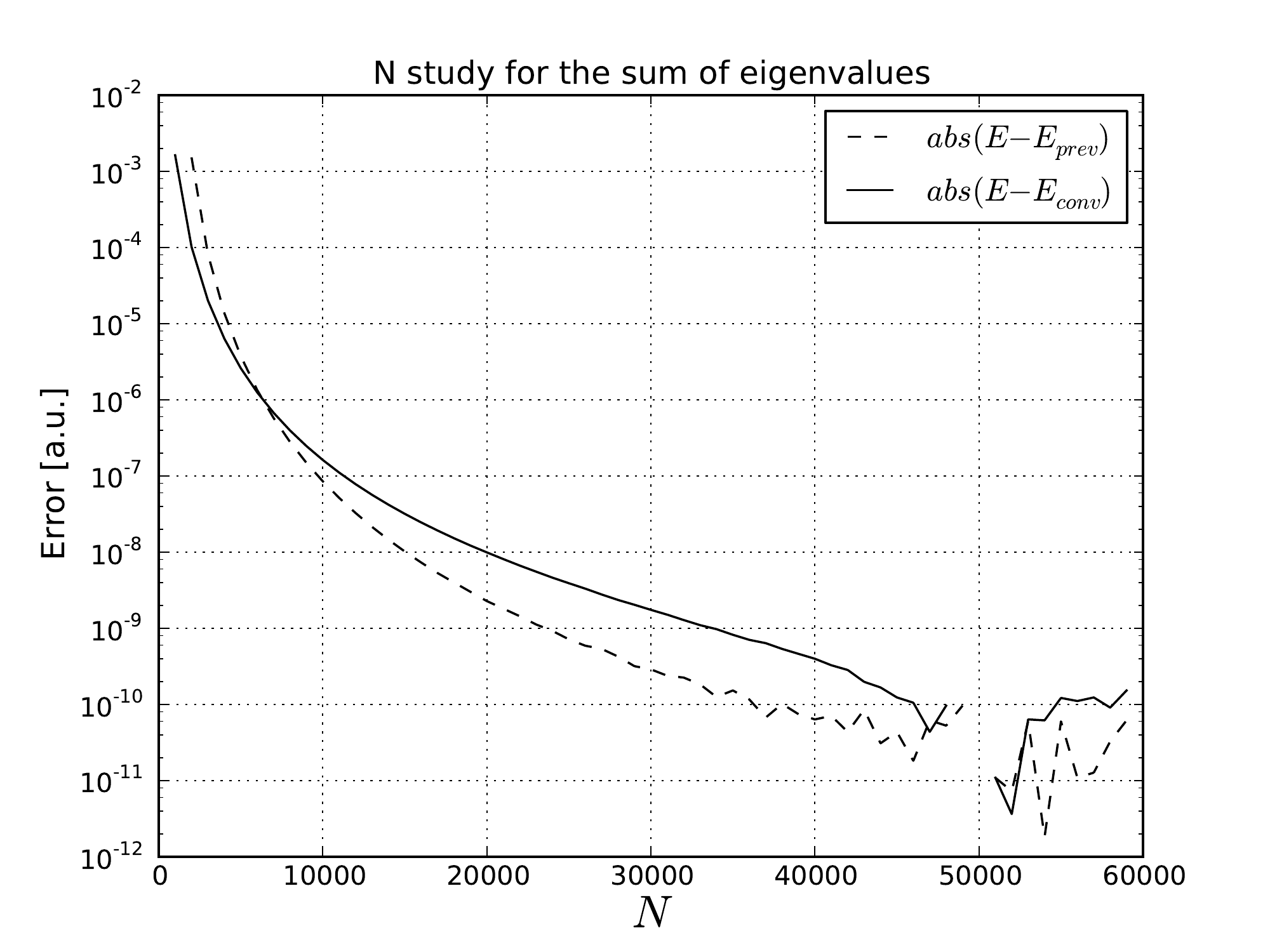}
}
\caption{Convergence studies for the sum of Schr\"odinger eigenvalues for $V=-Z/r$.}
\end{figure}

For the Schr\"odinger equation, 
the $r_\min$ convergence study was run with $a=10^9$, $r_\max=50$, 
and $N=50000$. From the resulting plot (Fig.~\ref{fig:Z_rmin}), we see that $r_\min=10^{-7}$ is a converged
value; decreasing $r_\min$ further does not reduce energy differences further. The $r_\max$ convergence study was run with $a=2.7 \cdot 10^6$, $r_\min=10^{-7}$, and $N=50000$.
From the resulting plot (Fig.~\ref{fig:Z_rmax}), we see that $r_\max=3.5$ is a converged value. Finally, the $N$
study was run with $a=2.7 \cdot 10^6$, $r_\min=10^{-7}$, and $r_\max=50$. From the
plot (Fig.~\ref{fig:Z_n_study}), it is determined that $N=50000$ is a converged value. 

From these results, it is
seen that the fixed parameters for each convergence study were fully
converged and the error indicator in each case was below $10^{-10}$ a.u. 
Table~\ref{table:Zr_eigs} compares the computed eigenvalues to exact 
values ($-{Z^2\over2n^2}$) for this analytic test case 
and it is verified that the absolute error of the computed values is below $10^{-10}$, 
as indicated by the convergence studies.

\begin{table}
\begin{center}
{\tiny
\begin{tabular}{c c c c c}
n & l & \ttt{dftatom} & exact & diff \\
\hline
1 & 0 & -4232.00000000001 & -4232.00000000000 & 9.09e-12 \\
2 & 0 & -1058.00000000001 & -1058.00000000000 & 7.28e-12 \\
2 & 1 & -1058.00000000000 & -1058.00000000000 & 9.09e-13 \\
3 & 0 & -470.222222222232 & -470.222222222222 & 9.55e-12 \\
3 & 1 & -470.222222222224 & -470.222222222222 & 2.22e-12 \\
3 & 2 & -470.222222222224 & -470.222222222222 & 2.22e-12 \\
4 & 0 & -264.500000000013 & -264.500000000000 & 1.33e-11 \\
4 & 1 & -264.500000000005 & -264.500000000000 & 5.00e-12 \\
4 & 2 & -264.500000000005 & -264.500000000000 & 5.00e-12 \\
4 & 3 & -264.499999999997 & -264.500000000000 & 3.18e-12 \\
5 & 0 & -169.280000000016 & -169.280000000000 & 1.61e-11 \\
5 & 1 & -169.280000000011 & -169.280000000000 & 1.08e-11 \\
5 & 2 & -169.280000000006 & -169.280000000000 & 5.57e-12 \\
5 & 3 & -169.280000000006 & -169.280000000000 & 5.57e-12 \\
5 & 4 & -169.280000000000 & -169.280000000000 & 3.13e-13 \\
6 & 0 & -117.555555555579 & -117.555555555556 & 2.34e-11 \\
6 & 1 & -117.555555555579 & -117.555555555556 & 2.34e-11 \\
6 & 2 & -117.555555555572 & -117.555555555556 & 1.61e-11 \\
6 & 3 & -117.555555555564 & -117.555555555556 & 8.84e-12 \\
6 & 4 & -117.555555555557 & -117.555555555556 & 1.55e-12 \\
6 & 5 & -117.555555555557 & -117.555555555556 & 1.55e-12 \\
7 & 0 & -86.3673469388100 & -86.3673469387755 & 3.44e-11 \\
7 & 1 & -86.3673469388046 & -86.3673469387755 & 2.91e-11 \\
7 & 2 & -86.3673469387993 & -86.3673469387755 & 2.38e-11 \\
7 & 3 & -86.3673469387886 & -86.3673469387755 & 1.31e-11 \\
7 & 4 & -86.3673469387832 & -86.3673469387755 & 7.73e-12 \\
7 & 5 & -86.3673469387779 & -86.3673469387755 & 2.36e-12 \\
7 & 6 & -86.3673469387779 & -86.3673469387755 & 2.36e-12 \\
\end{tabular}
}
\caption{Computed and exact Schr\"odinger eigenvalues for $V = -{Z\over r}$.}
\label{table:Zr_eigs}
\end{center}
\end{table}

Of course, smaller $N$ could be used to achieve the above accuracies 
using the converged $r_\min$ and $r_\max$ values indicated by the above studies, 
then optimizing the mesh gradation $a$ to minimize the required $N$. The purpose of the present 
analytic case study, however, is to verify the accuracy of the solver and establish 
the robustness of convergence indicators, $E-E_\mrm{prev}$ and $E-E_\mrm{conv}$. 
Optimal $a$ and $N$ are discussed in Section~\ref{sec:U}.

\begin{table}
\begin{center}
{\tiny
\begin{tabular}{c c c c c c}
n & l & s & \ttt{dftatom} & exact & diff \\
\hline
1 & 0 & 1 & -4861.19802311923 & -4861.19802311937 & 1.41e-10 \\
2 & 0 & 1 & -1257.39589025788 & -1257.39589025789 & 8.41e-12 \\
2 & 1 & 1 & -1089.61142091988 & -1089.61142091987 & 1.59e-12 \\
2 & 1 & -1 & -1257.39589025790 & -1257.39589025789 & 1.75e-11 \\
3 & 0 & 1 & -539.093341793886 & -539.093341793890 & 3.98e-12 \\
3 & 1 & 1 & -489.037087678200 & -489.037087678200 & 1.14e-13 \\
3 & 1 & -1 & -539.093341793896 & -539.093341793890 & 5.68e-12 \\
3 & 2 & 1 & -476.261595161156 & -476.261595161155 & 1.36e-12 \\
3 & 2 & -1 & -489.037087678199 & -489.037087678200 & 6.25e-13 \\
4 & 0 & 1 & -295.257844100398 & -295.257844100397 & 1.42e-12 \\
4 & 1 & 1 & -274.407758840063 & -274.407758840065 & 2.50e-12 \\
4 & 1 & -1 & -295.257844100398 & -295.257844100397 & 1.25e-12 \\
4 & 2 & 1 & -268.965877827131 & -268.965877827130 & 9.66e-13 \\
4 & 2 & -1 & -274.407758840063 & -274.407758840065 & 1.88e-12 \\
4 & 3 & 1 & -266.389447187815 & -266.389447187816 & 3.41e-13 \\
4 & 3 & -1 & -268.965877827131 & -268.965877827130 & 1.53e-12 \\
5 & 0 & 1 & -185.485191678549 & -185.485191678552 & 3.01e-12 \\
5 & 1 & 1 & -174.944613583463 & -174.944613583462 & 7.67e-13 \\
5 & 1 & -1 & -185.485191678550 & -185.485191678552 & 1.79e-12 \\
5 & 2 & 1 & -172.155252323734 & -172.155252323737 & 2.67e-12 \\
5 & 2 & -1 & -174.944613583463 & -174.944613583462 & 6.54e-13 \\
5 & 3 & 1 & -170.828937049879 & -170.828937049879 & 1.14e-13 \\
5 & 3 & -1 & -172.155252323735 & -172.155252323737 & 1.88e-12 \\
5 & 4 & 1 & -170.049934288550 & -170.049934288552 & 2.36e-12 \\
5 & 4 & -1 & -170.828937049878 & -170.828937049879 & 1.19e-12 \\
6 & 0 & 1 & -127.093638842628 & -127.093638842631 & 2.80e-12 \\
6 & 1 & 1 & -121.057538029547 & -121.057538029549 & 1.44e-12 \\
6 & 1 & -1 & -127.093638842628 & -127.093638842631 & 2.26e-12 \\
6 & 2 & 1 & -119.445271987140 & -119.445271987141 & 6.39e-13 \\
6 & 2 & -1 & -121.057538029547 & -121.057538029549 & 1.38e-12 \\
6 & 3 & 1 & -118.676410324351 & -118.676410324351 & 1.28e-13 \\
6 & 3 & -1 & -119.445271987140 & -119.445271987141 & 3.13e-13 \\
6 & 4 & 1 & -118.224144624903 & -118.224144624903 & 2.42e-13 \\
6 & 4 & -1 & -118.676410324351 & -118.676410324351 & 1.71e-13 \\
6 & 5 & 1 & -117.925825597294 & -117.925825597293 & 1.34e-12 \\
6 & 5 & -1 & -118.224144624903 & -118.224144624903 & 4.41e-13 \\
7 & 0 & 1 & -92.4407876009401 & -92.4407876009427 & 2.59e-12 \\
7 & 1 & 1 & -88.6717490520179 & -88.6717490520168 & 1.05e-12 \\
7 & 1 & -1 & -92.4407876009404 & -92.4407876009427 & 2.30e-12 \\
7 & 2 & 1 & -87.6582876318895 & -87.6582876318935 & 3.94e-12 \\
7 & 2 & -1 & -88.6717490520179 & -88.6717490520168 & 1.08e-12 \\
7 & 3 & 1 & -87.1739666719491 & -87.1739666719477 & 1.35e-12 \\
7 & 3 & -1 & -87.6582876318897 & -87.6582876318935 & 3.77e-12 \\
7 & 4 & 1 & -86.8887663909435 & -86.8887663909409 & 2.64e-12 \\
7 & 4 & -1 & -87.1739666719491 & -87.1739666719477 & 1.38e-12 \\
7 & 5 & 1 & -86.7005195728073 & -86.7005195728088 & 1.55e-12 \\
7 & 5 & -1 & -86.8887663909439 & -86.8887663909409 & 3.06e-12 \\
7 & 6 & 1 & -86.5668751023587 & -86.5668751023550 & 3.77e-12 \\
7 & 6 & -1 & -86.7005195728072 & -86.7005195728088 & 1.63e-12 \\
\end{tabular}
}
\caption{Computed and exact Dirac eigenvalues for $V = -{Z\over r}$.}
\label{table:Zr_eigs_rel}
\end{center}
\end{table}

For the Dirac equation, the $N$ study is first done for $r_\min$ larger than
$10^{-11}$ and $N=200000$ was obtained as a converged value ($a=10^9$ was used so
that convergence was obtained for tractable $N$ for all $r_\min$). 
The $r_\min$ study was then run and determined $r_\min=10^{-8}$ as a converged value, 
giving accuracy better than $5 \cdot 10^{-10}\rm\,a.u.$ $r_\max=50$ was used
and it was verified that it is a converged value.
Table~\ref{table:Zr_eigs_rel} shows a comparison of computed and exact
Dirac eigenvalues
\begin{align}
E_{n\kappa} & = {c^2\over\sqrt{1 + {(Z/c)^2\over (n - |\kappa| + \beta)^2}}} - c^2, \\
\beta       & = \sqrt{\kappa^2 - \left(Z/c\right)^2},
\end{align}
for the Coulomb problem, and it is verified that
the absolute error is below $5 \cdot 10^{-10}\rm\,a.u.$, 
as indicated by the convergence studies.
Note: all numerical and analytic calculations in this paper use the
1986 CODATA \cite{Cohen1987} recommended value
$c=137.0359895\rm\, a.u.$, the same value as in NIST benchmarks
\cite{nist}.

\subsection{Nonsingular test case: $V = \half \omega^2 r^2$}

Table \ref{table:osc_eigs} shows computed Schr\"odinger eigenvalues for the harmonic oscillator potential
$V(r) = \half \omega^2 r^2$ with $\omega = 1$ compared to exact values
\begin{equation}
E_{nl} = \omega\left(2n-l-{1\over2}\right).
\end{equation}
Asymptotics \eqref{sch_asympt_outward}, \eqref{sch_asympt_outward2} are used for outward integration, with perturbation correction turned
off due to positive energies. It is seen that $r_\min=10^{-7}\rm\,a.u.$, $r_\max=10\rm\,a.u.$, $a=10$ and $N=5000$ are sufficient to obtain eigenvalues accurate to $< 10^{-8}$ a.u.

\begin{table}
\begin{center}
{\tiny
\begin{tabular}{c c c c c}
n & l & \ttt{dftatom} & exact & diff \\
\hline
  1 &  0 &   1.500000000168 &   1.500000000000 &  1.68E-10 \\
  2 &  0 &   3.500000000587 &   3.500000000000 &  5.87E-10 \\
  2 &  1 &   2.499999999999 &   2.500000000000 &  8.53E-13 \\
  3 &  0 &   5.500000001140 &   5.500000000000 &  1.14E-09 \\
  3 &  1 &   4.499999999993 &   4.500000000000 &  7.22E-12 \\
  3 &  2 &   3.499999999999 &   3.500000000000 &  1.19E-12 \\
  4 &  0 &   7.500000001777 &   7.500000000000 &  1.78E-09 \\
  4 &  1 &   6.499999999968 &   6.500000000000 &  3.18E-11 \\
  4 &  2 &   5.499999999990 &   5.500000000000 &  1.04E-11 \\
  4 &  3 &   4.499999999998 &   4.500000000000 &  1.53E-12 \\
  5 &  0 &   9.500000002451 &   9.500000000000 &  2.45E-09 \\
  5 &  1 &   8.499999999900 &   8.500000000000 &  1.00E-10 \\
  5 &  2 &   7.499999999955 &   7.500000000000 &  4.52E-11 \\
  5 &  3 &   6.499999999986 &   6.500000000000 &  1.42E-11 \\
  5 &  4 &   5.499999999998 &   5.500000000000 &  2.44E-12 \\
  6 &  0 &  11.500000003104 &  11.500000000000 &  3.10E-09 \\
  6 &  1 &  10.499999999751 &  10.500000000000 &  2.49E-10 \\
  6 &  2 &   9.499999999864 &   9.500000000000 &  1.36E-10 \\
  6 &  3 &   8.499999999940 &   8.500000000000 &  6.03E-11 \\
  6 &  4 &   7.499999999982 &   7.500000000000 &  1.85E-11 \\
  6 &  5 &   6.499999999997 &   6.500000000000 &  2.79E-12 \\
  7 &  0 &  13.500000003665 &  13.500000000000 &  3.66E-09 \\
  7 &  1 &  12.499999999466 &  12.500000000000 &  5.34E-10 \\
  7 &  2 &  11.499999999672 &  11.500000000000 &  3.28E-10 \\
  7 &  3 &  10.499999999824 &  10.500000000000 &  1.76E-10 \\
  7 &  4 &   9.499999999923 &   9.500000000000 &  7.66E-11 \\
  7 &  5 &   8.499999999977 &   8.500000000000 &  2.34E-11 \\
  7 &  6 &   7.499999999996 &   7.500000000000 &  3.69E-12 \\
\end{tabular}
}
\caption{Computed and exact Schr\"odinger eigenvalues for $V = \half r^2$.}
\label{table:osc_eigs}
\end{center}
\end{table}

Table \ref{table:osc_eigs_rel} shows Dirac eigenvalues for the harmonic
oscillator potential 
$V(r) = \half \omega^2 r^2$ with $\omega=1$. 
Asymptotics \eqref{dirac_asympt_outward_nonsing1},
\eqref{dirac_asympt_outward_nonsing1b} and
\eqref{dirac_asympt_outward_nonsing2}, \eqref{dirac_asympt_outward_nonsing2b}
are used for outward integration, with perturbation correction again turned off
due to positive energies. Lacking analytic results for comparison, we compared
instead to independent high-order finite-element based calculations
\cite{CerP12} which require no assumptions of asymptotic forms or determination
of sufficiently small $r_\min>0$. For $r_\min=10^{-8}\rm\,a.u.$,
$r_\max=10\rm\,a.u.$, $a=80$, and $N=5000$, we find agreement of all eigenvalues
to $< 10^{-8}$ a.u., with values very near to the corresponding Schr\"odinger
results due to the relative weakness of the potential, in stark contrast to the
singular Coulombic case. Note that, unlike the case of a singular potential,
the number of nodes for $\kappa>0$ is only $n-l$, whereas the number of nodes
for $\kappa<0$ remains $n-l-1$, as in the singular case.

\begin{table}
\begin{center}
{\tiny
\begin{tabular}{c c c c}
n & l & s & \ttt{dftatom} \\
\hline
  1 &  0 &  0 &       1.49999501 \\
  2 &  0 &  0 &       3.49989517 \\
  2 &  1 &  0 &       2.49997504 \\
  2 &  1 &  1 &       2.49993510 \\
  3 &  0 &  0 &       5.49971548 \\
  3 &  1 &  0 &       4.49983527 \\
  3 &  1 &  1 &       4.49979534 \\
  3 &  2 &  0 &       3.49994176 \\
  3 &  2 &  1 &       3.49987520 \\
  4 &  0 &  0 &       7.49945594 \\
  4 &  1 &  0 &       6.49961565 \\
  4 &  1 &  1 &       6.49957572 \\
  4 &  2 &  0 &       5.49976206 \\
  4 &  2 &  1 &       5.49969551 \\
  4 &  3 &  0 &       4.49989517 \\
  4 &  3 &  1 &       4.49980199 \\
  5 &  0 &  0 &       9.49911657 \\
  5 &  1 &  0 &       8.49931620 \\
  5 &  1 &  1 &       8.49927627 \\
  5 &  2 &  0 &       7.49950252 \\
  5 &  2 &  1 &       7.49943598 \\
  5 &  3 &  0 &       6.49967554 \\
  5 &  3 &  1 &       6.49958238 \\
  5 &  4 &  0 &       5.49983526 \\
  5 &  4 &  1 &       5.49971547 \\
\end{tabular}
\hspace{1em}
\begin{tabular}{c c c c}
n & l & s & \ttt{dftatom} \\
\hline
  6 &  0 &  0 &      11.49869739 \\
  6 &  1 &  0 &      10.49893692 \\
  6 &  1 &  1 &      10.49889699 \\
  6 &  2 &  0 &       9.49916315 \\
  6 &  2 &  1 &       9.49909661 \\
  6 &  3 &  0 &       8.49937608 \\
  6 &  3 &  1 &       8.49928292 \\
  6 &  4 &  0 &       7.49957572 \\
  6 &  4 &  1 &       7.49945594 \\
  6 &  5 &  0 &       6.49976205 \\
  6 &  5 &  1 &       6.49961565 \\
  7 &  0 &  0 &      13.49819839 \\
  7 &  1 &  0 &      12.49847782 \\
  7 &  1 &  1 &      12.49843790 \\
  7 &  2 &  0 &      11.49874396 \\
  7 &  2 &  1 &      11.49867742 \\
  7 &  3 &  0 &      10.49899680 \\
  7 &  3 &  1 &      10.49890364 \\
  7 &  4 &  0 &       9.49923634 \\
  7 &  4 &  1 &       9.49911657 \\
  7 &  5 &  0 &       8.49946258 \\
  7 &  5 &  1 &       8.49931619 \\
  7 &  6 &  0 &       7.49967553 \\
  7 &  6 &  1 &       7.49950251 \\
\end{tabular}
}
\caption{Computed Dirac eigenvalues for $V = \half r^2$.}
\label{table:osc_eigs_rel}
\end{center}
\end{table}

\subsection{Double minimum potential}

A problem that is known to cause difficulties for shooting solvers is the
double minimum problem. A classical example is
the electronically excited $E$, $F\,{}^1\Sigma_g^+$ state of the hydrogen
molecule \cite{Kolos1969,Lin1974,Tobin1975}, but such potentials occur in electronic structure also \cite{Connerade1998}.

\begin{figure}
\centering
\includegraphics[width=0.5\linewidth]{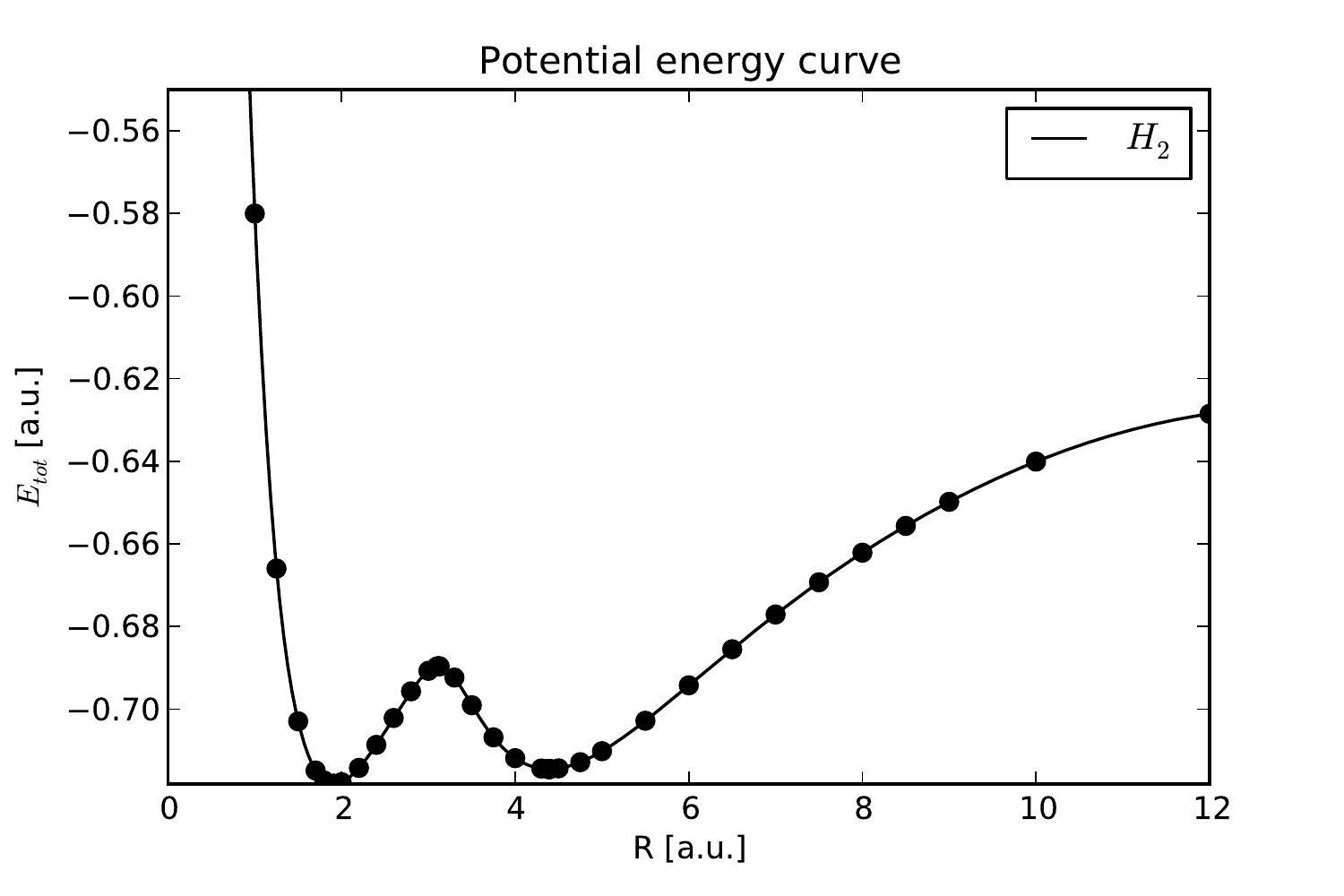}
\caption{Double minimum potential discrete values (dots) and cubic spline
interpolation (line).}
\label{fig:double_min_potential}
\end{figure}

The potential is given on a discrete grid in Table I of Ref.~\cite{Kolos1969}.
There are several options for interpolating the potential, each giving slightly
different energies. For the present purposes, cubic
spline interpolation was used, with consistent boundary conditions at
both ends (second derivative consistent with cubic interpolant), see
Fig.~\ref{fig:double_min_potential}. We have implemented Lagrange (as in \cite{Lin1974}) and Hermite (as in \cite{Kolos1969}) interpolation also, included in the distribution of \ttt{dftatom} \cite{dftatom}.

\begin{figure}
\centering
\subfloat[$N$ study eigenvalues]{
\label{fig:double_min_N_prev}
\includegraphics[width=0.5\linewidth]{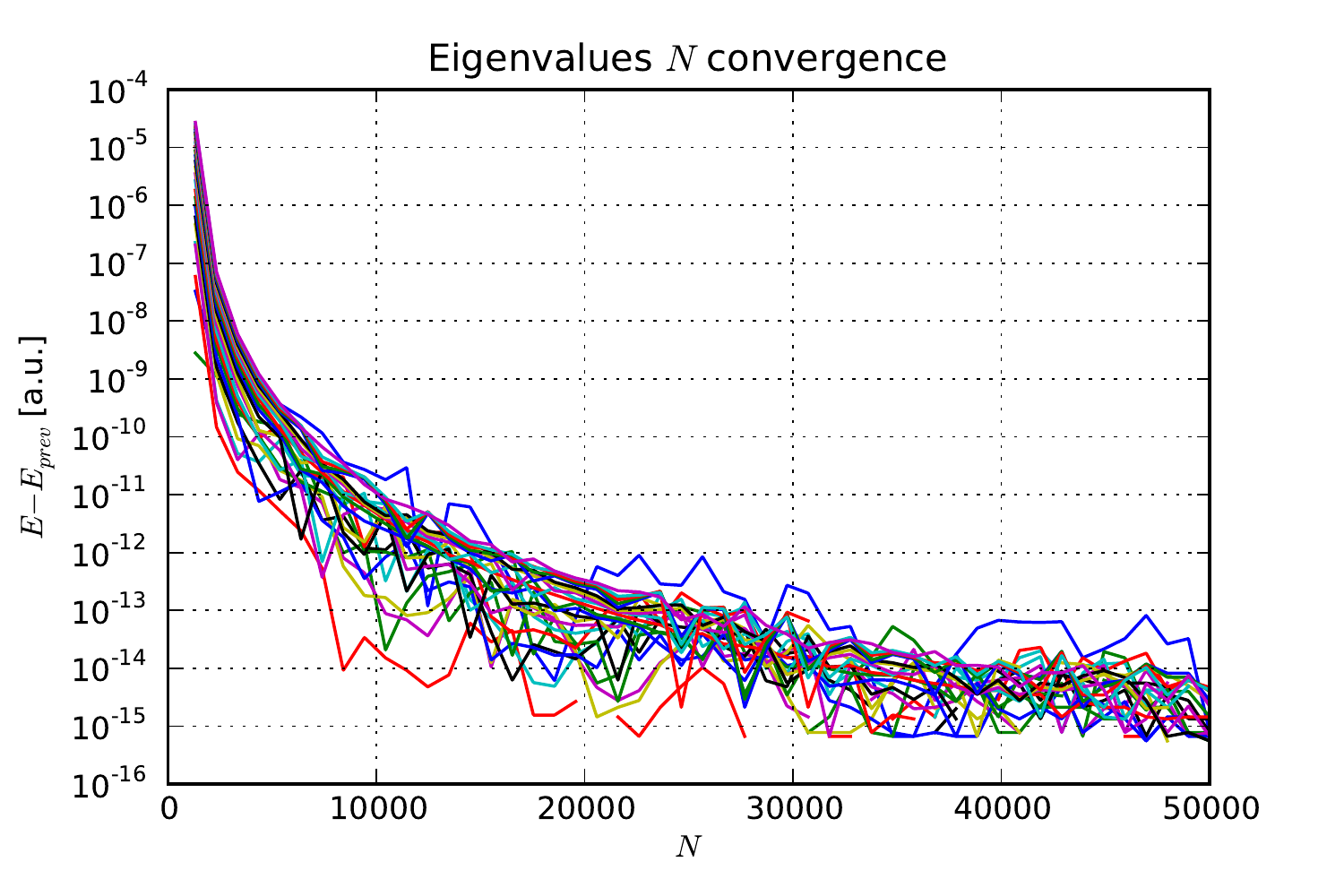}
}
\subfloat[$N$ study eigenvalues]{
\label{fig:double_min_N}
\includegraphics[width=0.5\linewidth]{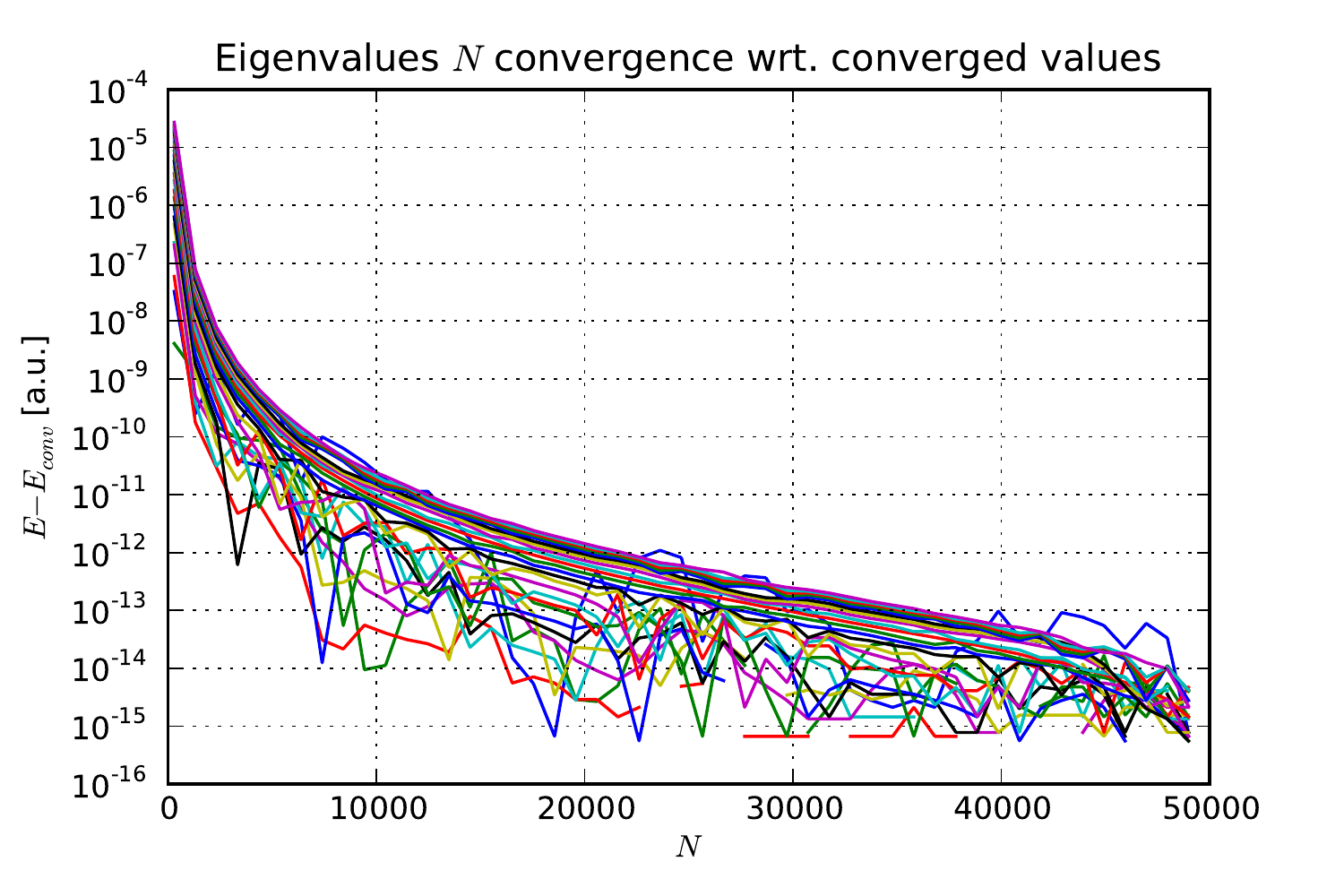}
}

\subfloat[$r_\min$ study eigenvalues]{
\label{fig:double_min_prev}
\includegraphics[width=0.5\linewidth]{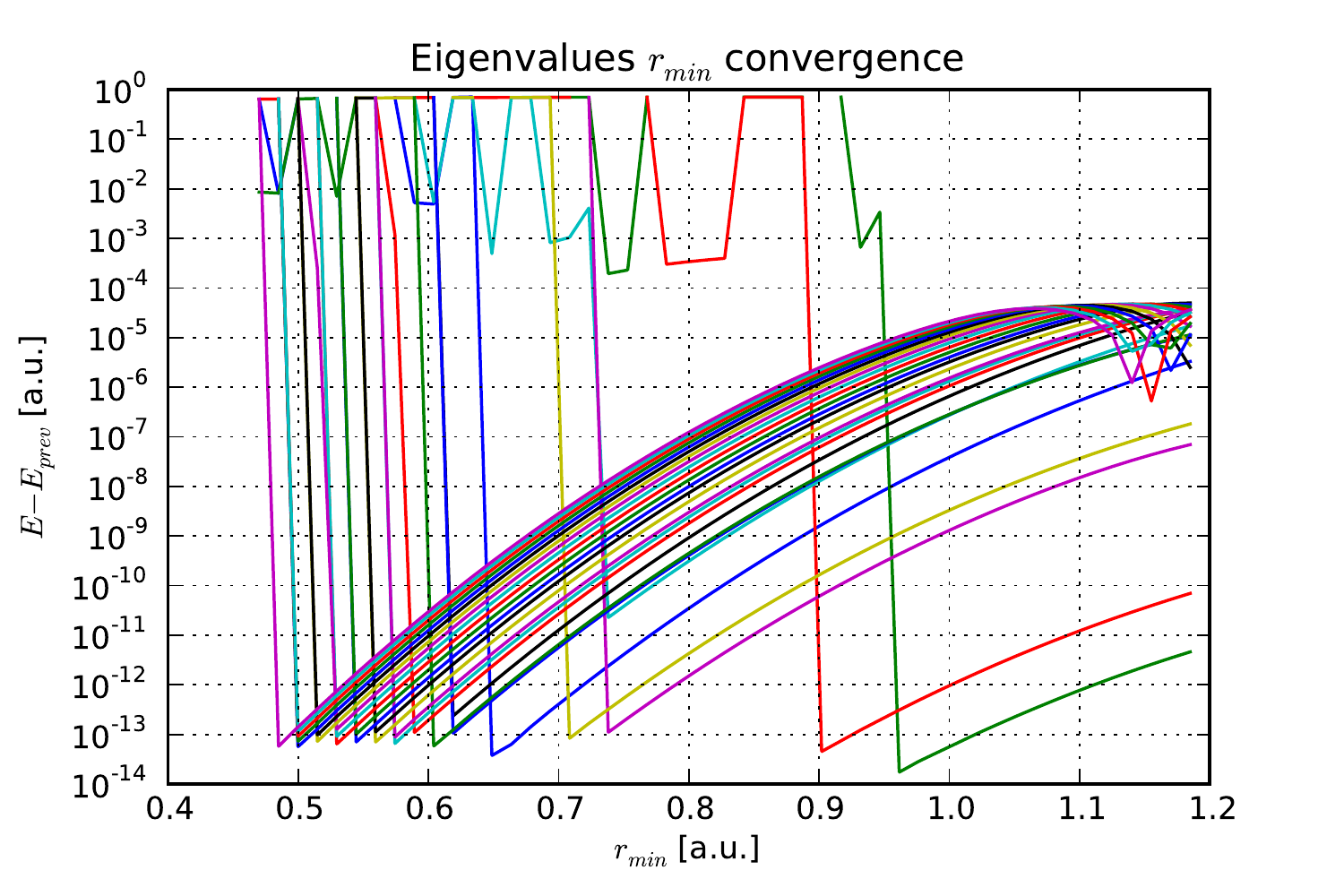}
}
\subfloat[$r_\min$ study eigenvalues]{
\label{fig:double_min}
\includegraphics[width=0.5\linewidth]{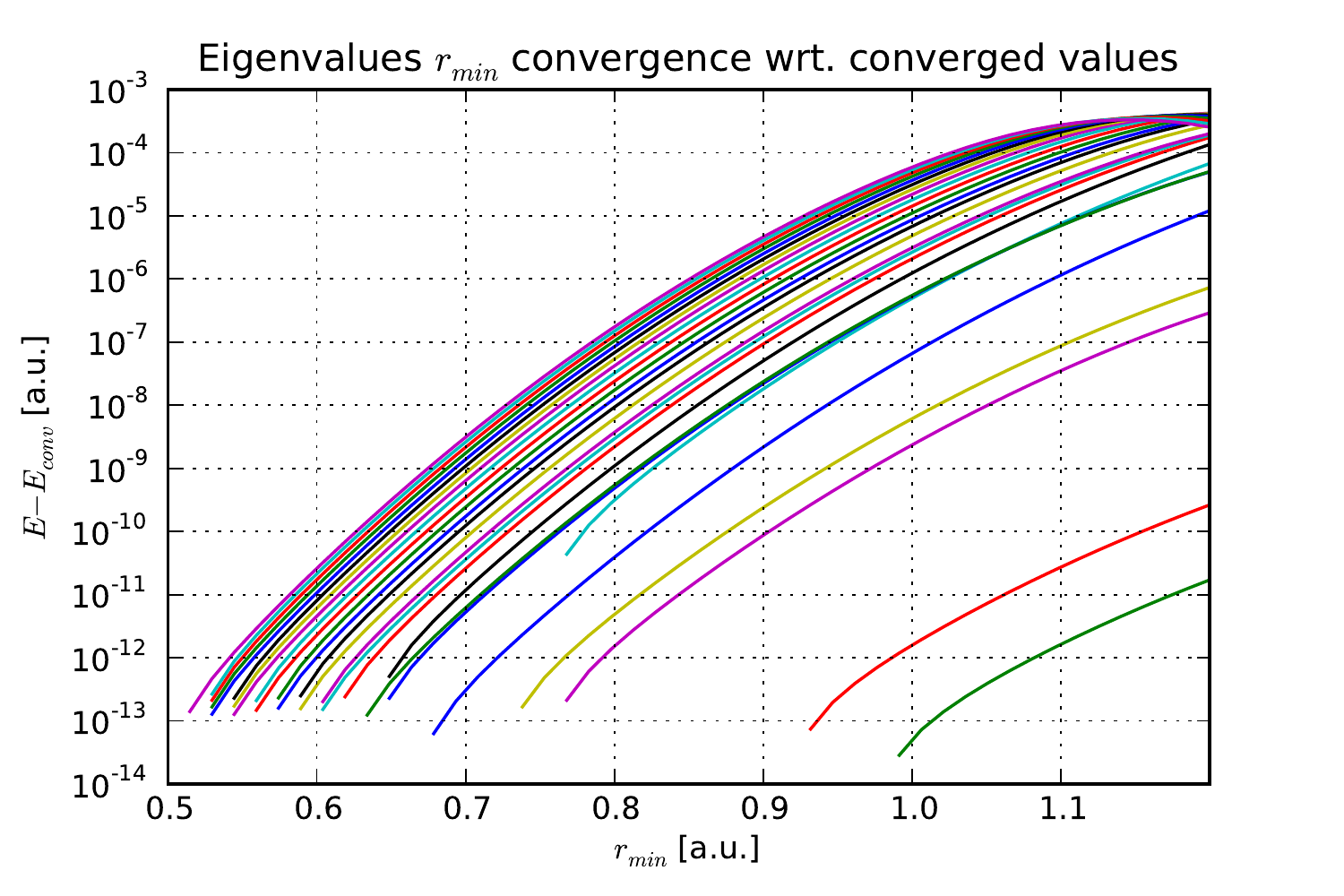}
}
\caption{Double minimum potential convergence studies.}
\end{figure}

The problem is solved for $r_\max=12$ a.u., with reduced mass $\mu = 1836.12/2$
a.u. and conversion of $E$ from a.u.\ to $\rm cm^{-1}$ via 
$(-0.625 - E) \cdot 219474.62$, as in \cite{Kolos1969}. 
First, an $N$-convergence study was
performed to determine $N=50000$ sufficient to converge all eigenvalues to 
$<10^{-12}$ a.u., see Figs.~\ref{fig:double_min_N_prev} and
\ref{fig:double_min_N}. Convergence with respect to $r_\min$ was then studied
with $N=50000$, see Fig.~\ref{fig:double_min_prev}, from which a converged energy and 
$r_\min$ were determined for each eigenvalue, as shown in Table~\ref{table:double_min}.
Fig.~\ref{fig:double_min} shows the convergence with respect to these converged
energies. From these, we find eigenvalues stable to $< 10^{-8}$ a.u. with
respect to both $r_\min$ and $N$.

To verify the results, 
we compared to independent high-order finite-element based
calculations \cite{CerP12} on the interval $[0, 12]$ a.u., which require no
assumptions of asymptotic forms or determination of $r_\min$. 
Agreement was obtained for all eigenvalues to 
$< 10^{-8}$ a.u. ($< 10^{-2}\rm\,cm^{-1}$). 
To compare to \cite{Tobin1975}, we implemented the finite difference method
described there and used a uniform mesh with $N=100000$ points on the interval
$[0, 12]$ a.u. The results are shown in the Table \ref{table:double_min}.  
As can be seen, agreement is again obtained to $< 10^{-8}$ a.u. ($<
10^{-2}\rm\,cm^{-1}$). We thus obtain agreement of all eigenvalues to $<
10^{-8}$ a.u. by three independent methods and codes. Remaining differences
from \cite{Kolos1969} and \cite{Tobin1975} are thus most likely attributable to
differences in interpolation, choice of $r_\min$, and/or convergence with
respect to other computational parameters.

\begin{table}
\begin{center}
{\tiny
\begin{tabular}{c r r r r r}
$v$ & $r_\min$ [a.u.] & $E$ [a.u.] (\ttt{dftatom}) & $E$ [a.u.] (FD) & $E$ [$\rm cm^{-1}$] (\ttt{dftatom}) & $E$ [$\rm cm^{-1}$] (FD) \\
\hline
 0 & 0.679 & -0.71248345 & -0.71248345 & 19200.40 & 19200.40 \\
 1 & 0.991 & -0.71169411 & -0.71169411 & 19027.16 & 19027.16 \\
 2 & 0.932 & -0.70624880 & -0.70624880 & 17832.05 & 17832.05 \\
 3 & 0.768 & -0.70187763 & -0.70187763 & 16872.69 & 16872.69 \\
 4 & 0.768 & -0.70106161 & -0.70106161 & 16693.59 & 16693.59 \\
 5 & 0.738 & -0.69617697 & -0.69617697 & 15621.54 & 15621.54 \\
 6 & 0.649 & -0.69265160 & -0.69265160 & 14847.81 & 14847.81 \\
 7 & 0.649 & -0.69127773 & -0.69127773 & 14546.28 & 14546.28 \\
 8 & 0.634 & -0.68736218 & -0.68736218 & 13686.92 & 13686.92 \\
 9 & 0.619 & -0.68429008 & -0.68429008 & 13012.67 & 13012.67 \\
10 & 0.604 & -0.68145547 & -0.68145547 & 12390.54 & 12390.54 \\
11 & 0.604 & -0.67811006 & -0.67811006 & 11656.31 & 11656.31 \\
12 & 0.589 & -0.67481518 & -0.67481519 & 10933.17 & 10933.17 \\
13 & 0.589 & -0.67163329 & -0.67163330 & 10234.82 & 10234.82 \\
14 & 0.574 & -0.66844493 & -0.66844493 &  9535.06 &  9535.06 \\
15 & 0.574 & -0.66526724 & -0.66526725 &  8837.64 &  8837.64 \\
16 & 0.559 & -0.66215075 & -0.66215075 &  8153.65 &  8153.65 \\
17 & 0.559 & -0.65910546 & -0.65910546 &  7485.28 &  7485.28 \\
18 & 0.544 & -0.65612571 & -0.65612571 &  6831.30 &  6831.30 \\
19 & 0.544 & -0.65321351 & -0.65321351 &  6192.15 &  6192.15 \\
20 & 0.544 & -0.65037671 & -0.65037671 &  5569.54 &  5569.54 \\
21 & 0.530 & -0.64762174 & -0.64762175 &  4964.90 &  4964.90 \\
22 & 0.530 & -0.64495237 & -0.64495238 &  4379.04 &  4379.04 \\
23 & 0.530 & -0.64237186 & -0.64237186 &  3812.68 &  3812.68 \\
24 & 0.530 & -0.63988486 & -0.63988486 &  3266.85 &  3266.85 \\
25 & 0.515 & -0.63749749 & -0.63749749 &  2742.88 &  2742.88 \\
\end{tabular}
}
\caption{Computed Schr\"odinger eigenvalues for double minimum potential.
The $r_\min$ column shows values that were used for the shooting method.
The finite difference (FD) method was used on the interval $[0, 12]$.
}
\label{table:double_min}
\end{center}
\end{table}

\subsection{Uranium}
\label{sec:U}

We now consider full, self-consistent nonrelativistic (LDA) and relativistic
(RLDA) Kohn--Sham calculations of the ground state of uranium: a stringent test case, as numerous eigenstates are required, with tightly bound, highly oscillatory $s$ states, spanning energies from $\sim -0.1$ to $\sim -4,000$~a.u.

We use the same electronic configuration as in \cite{nist,nistweb}: for
the nonrelativistic calculation, $1s^2$, $2s^2$, $2p^6$, $3s^2$, $3p^6$,
$3d^{10}$, $4s^2$, $4p^6$, $4d^{10}$, $4f^{14}$, $5s^2$, $5p^6$, $5d^{10}$,
$5f^3$, $6s^2$, $6p^6$, $6d^1$, $7s^2$; and for the relativistic calculation,
the occupation of each $l$-shell is split according to the
degeneracy of $j=l+\half$ and $j=l-\half$ subshells.

\begin{figure}
\centering
\subfloat[$r_\min$ study total energy]{
\label{fig:lda_rmin}
\includegraphics[width=0.5\linewidth]{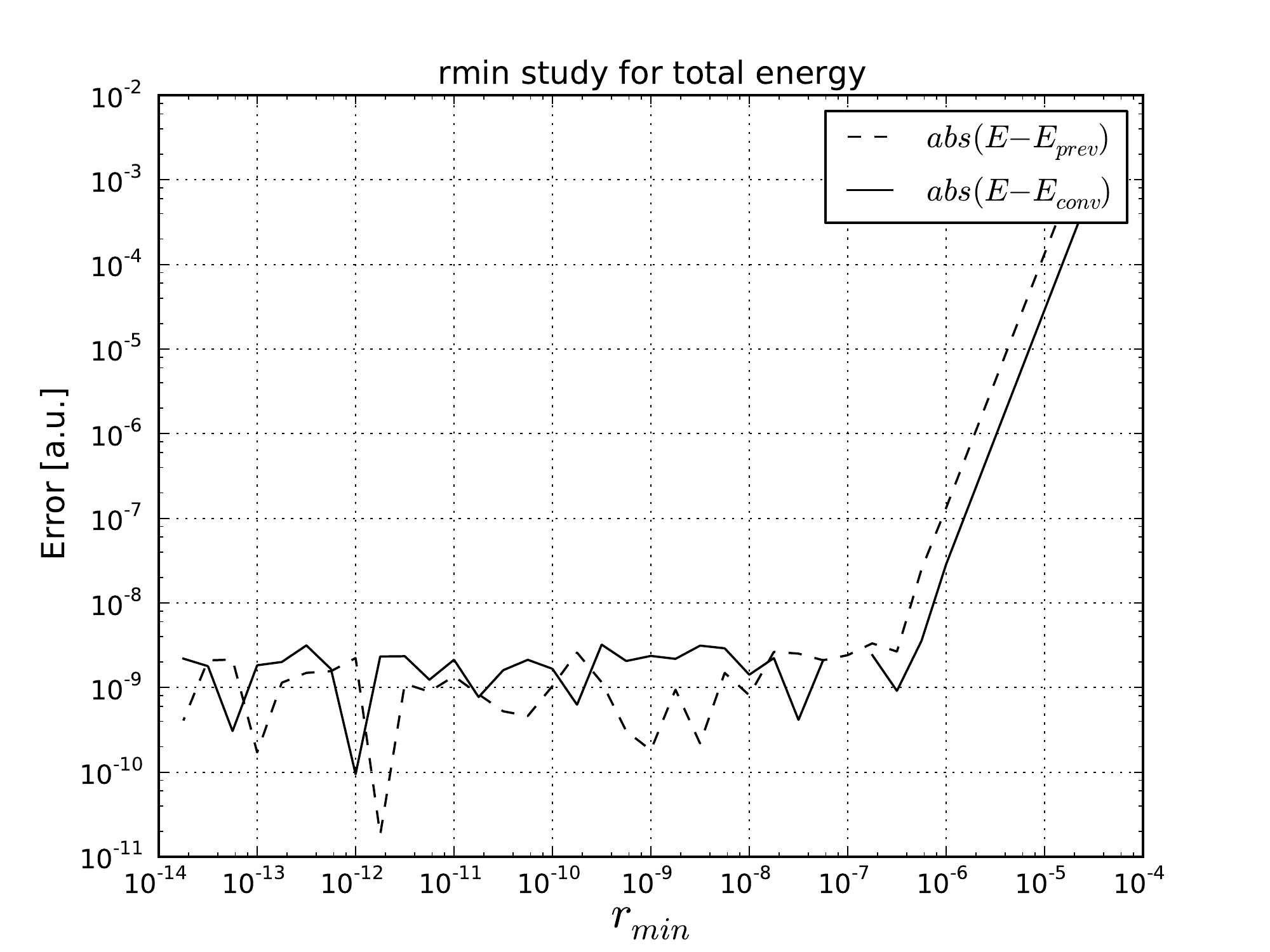}
}
\subfloat[$r_\max$ study eigenvalues]{
\label{fig:lda_rmax_eig}
\includegraphics[width=0.5\linewidth]{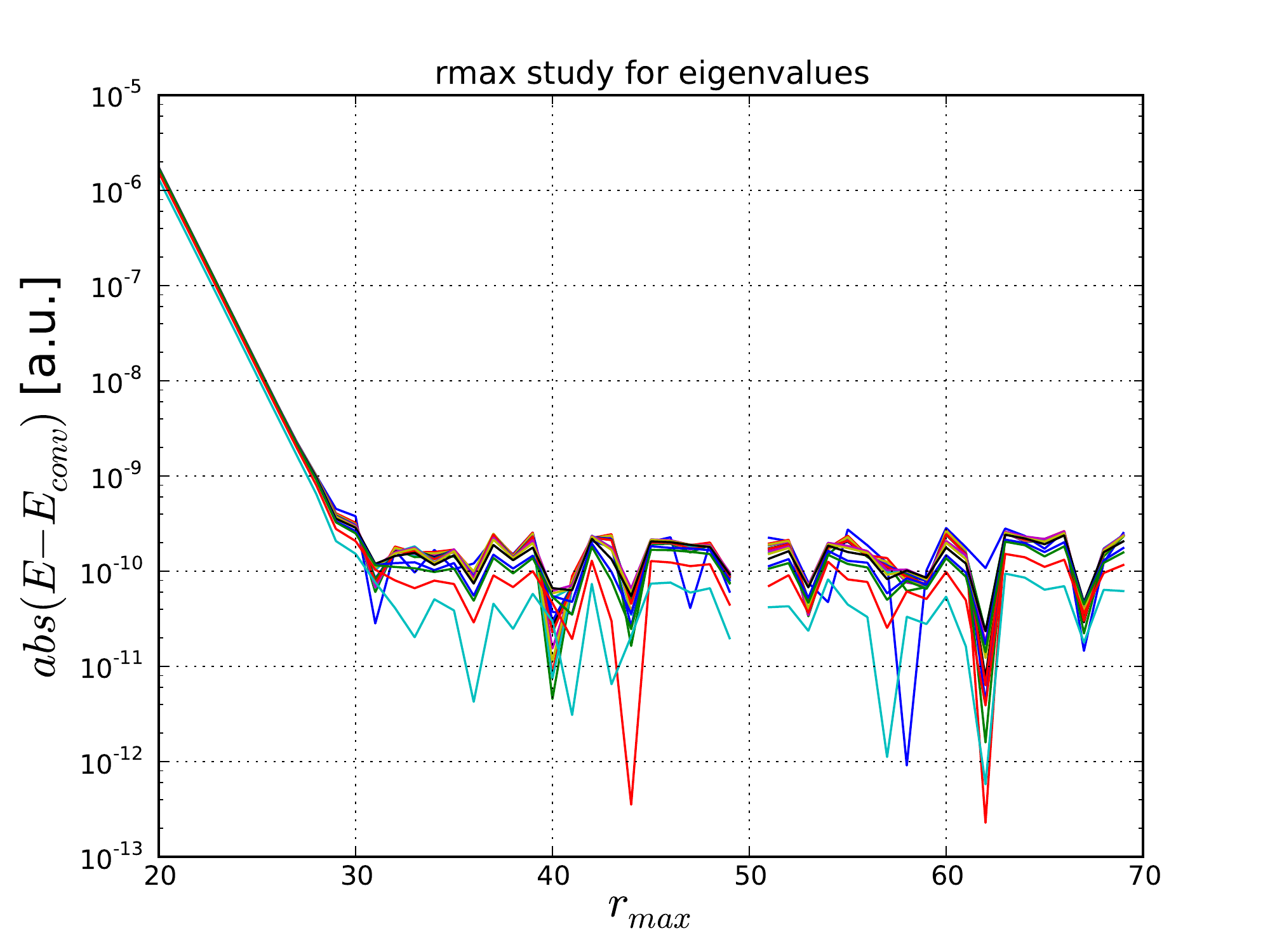}
}

\subfloat[$r_\max$ study total energy]{
\label{fig:lda_rmax_etot}
\includegraphics[width=0.5\linewidth]{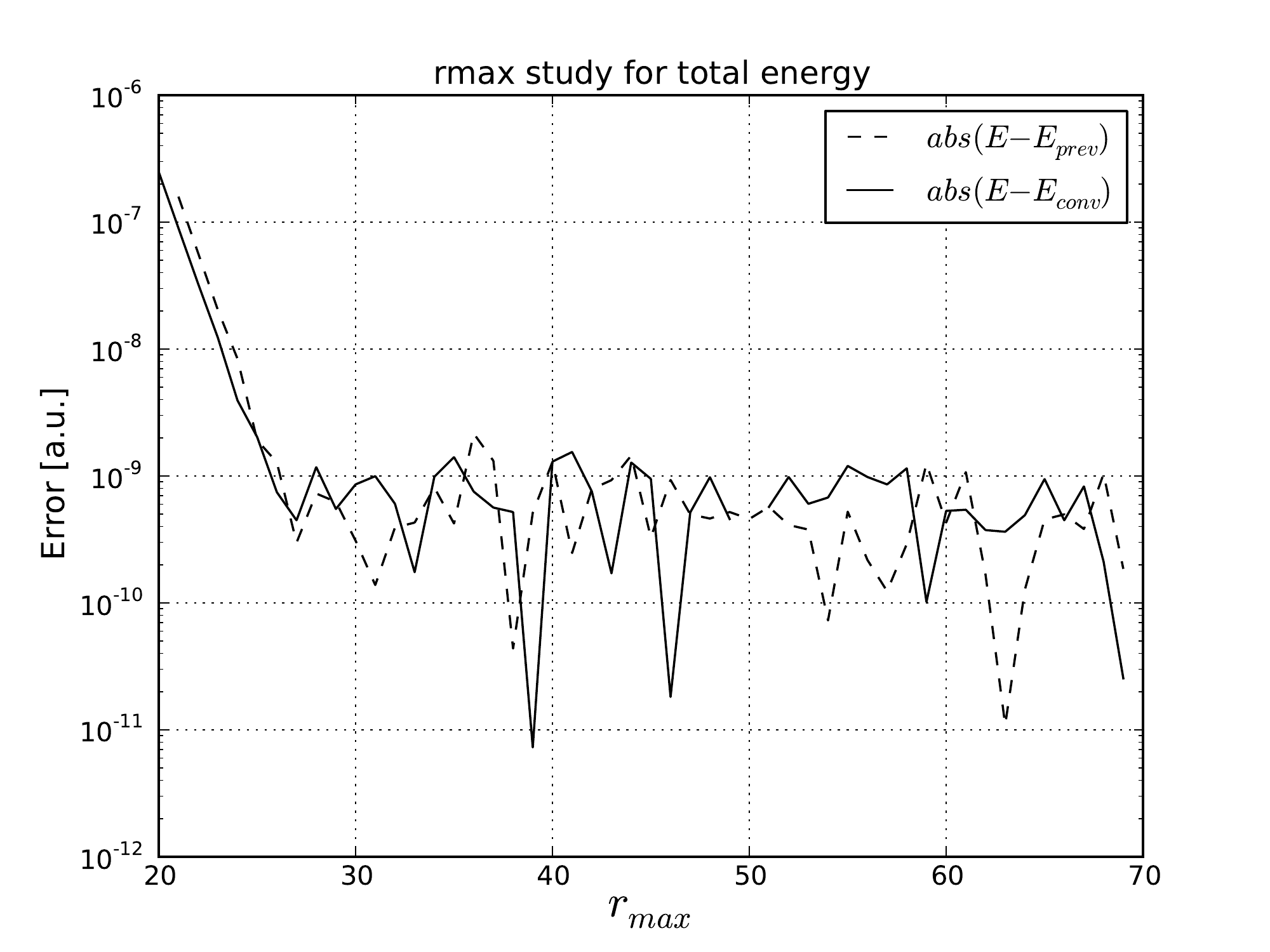}
}
\subfloat[$N$ study total energy]{
\label{fig:lda_N}
\includegraphics[width=0.5\linewidth]{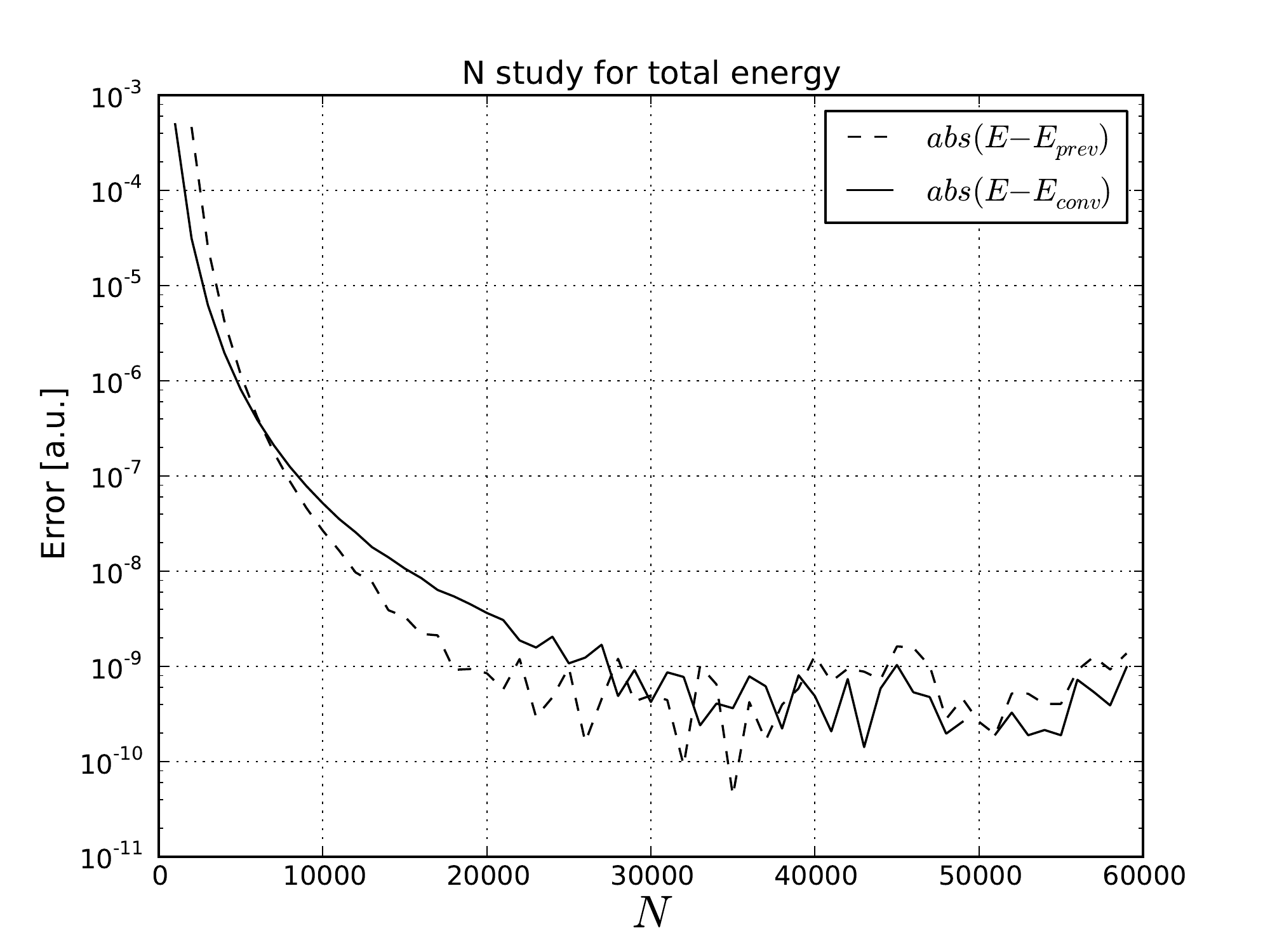}
}
\caption{Non-relativistic (LDA) density functional theory (DFT) convergence
studies for uranium.}
\end{figure}

\begin{figure}
\centering
\subfloat[$r_\min$ study total energy]{
\label{fig:rlda_rmin}
\includegraphics[width=0.5\linewidth]{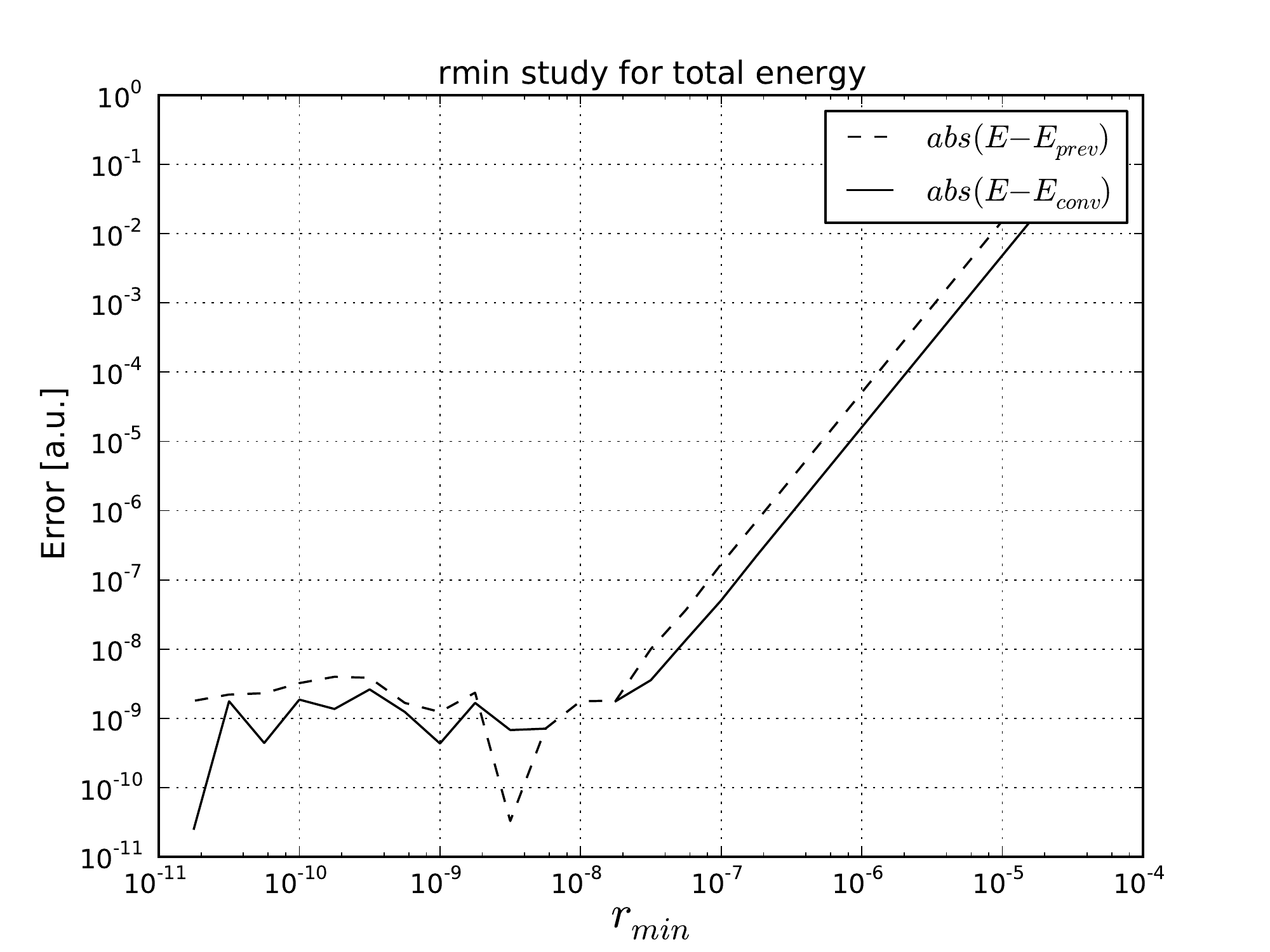}
}
\subfloat[$r_\max$ study eigenvalues]{
\label{fig:rlda_rmax_eig}
\includegraphics[width=0.5\linewidth]{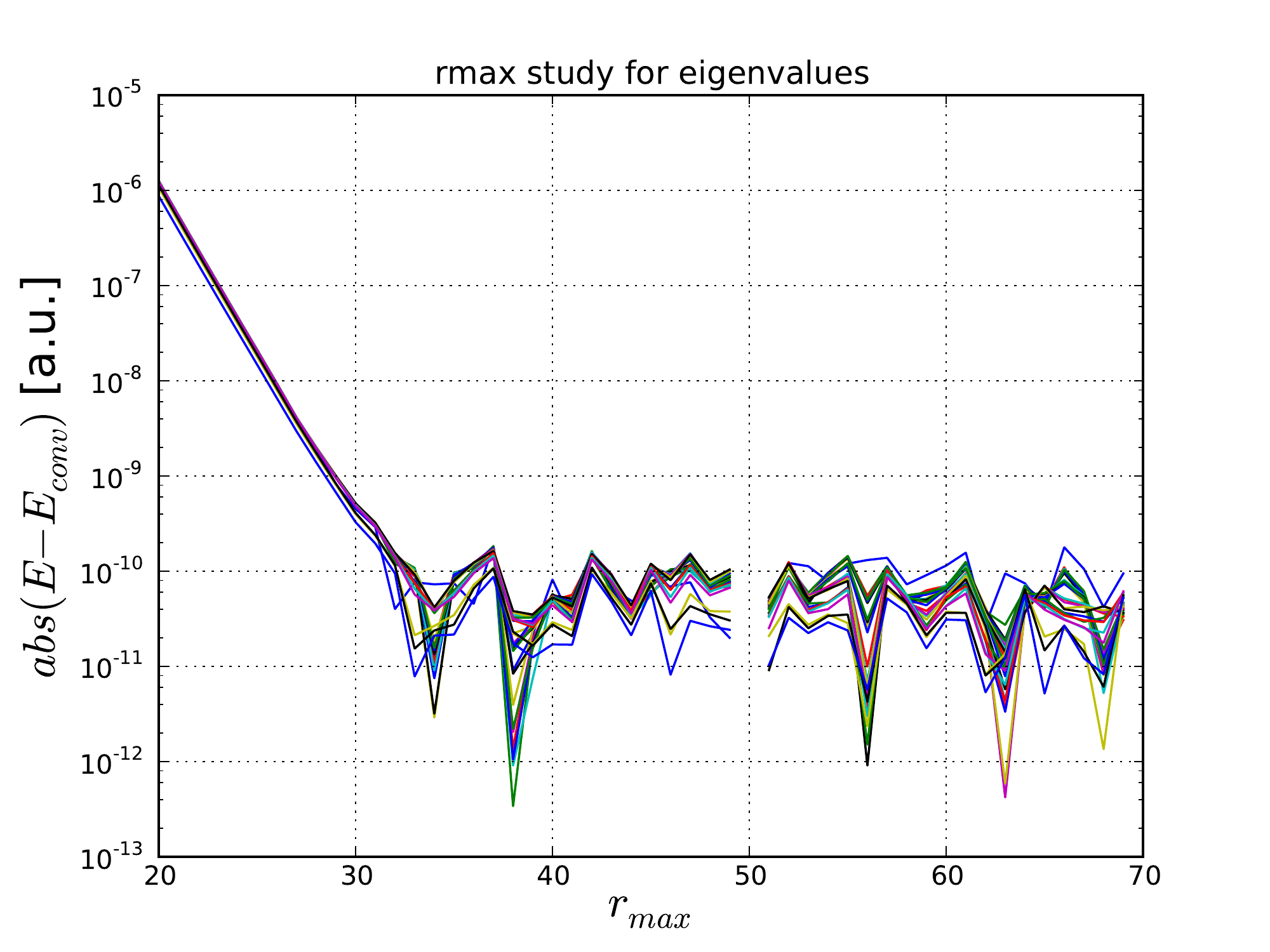}
}

\subfloat[$r_\max$ study total energy]{
\label{fig:rlda_rmax_etot}
\includegraphics[width=0.5\linewidth]{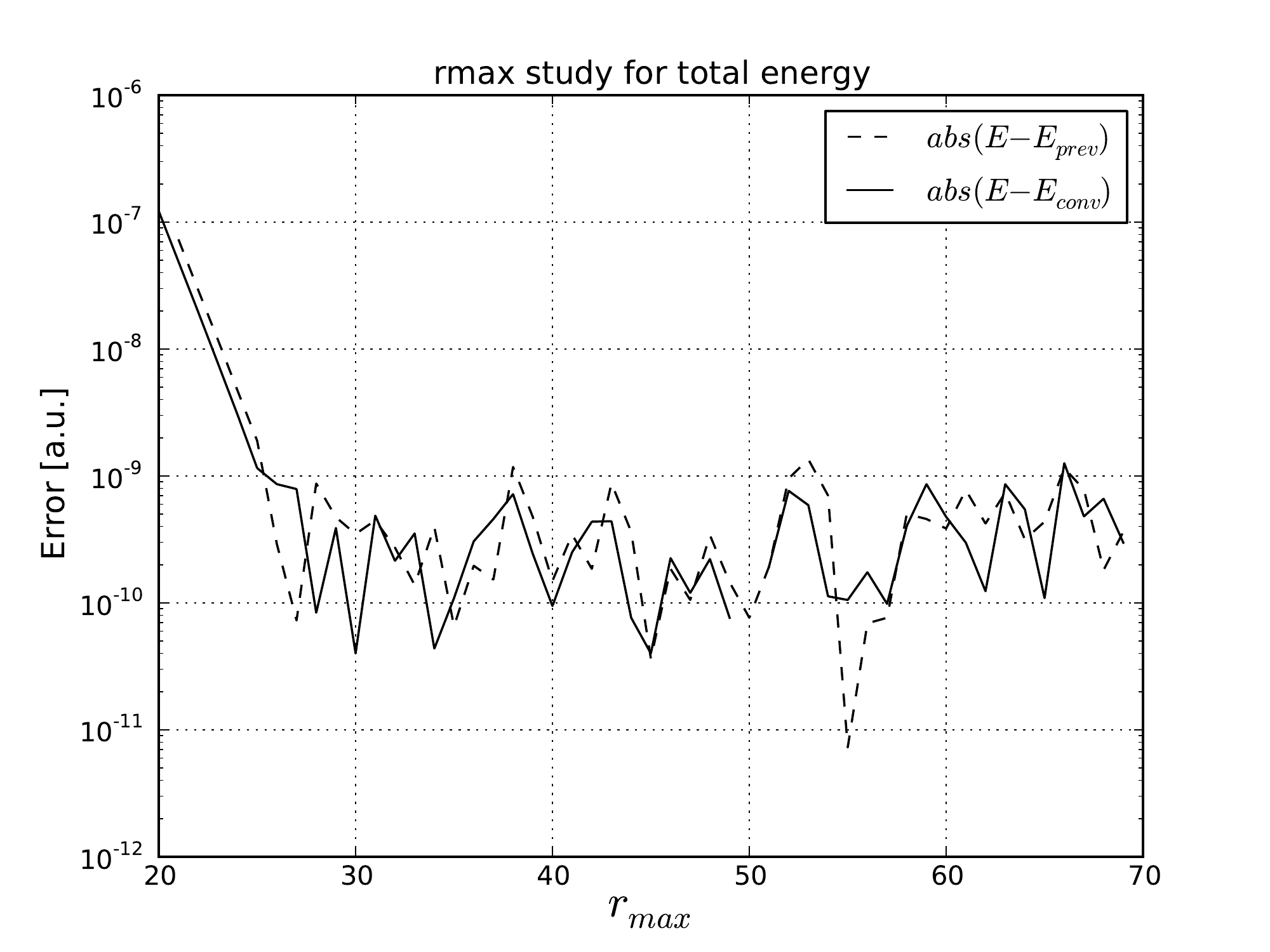}
}
\subfloat[$N$ study total energy]{
\label{fig:rlda_N}
\includegraphics[width=0.5\linewidth]{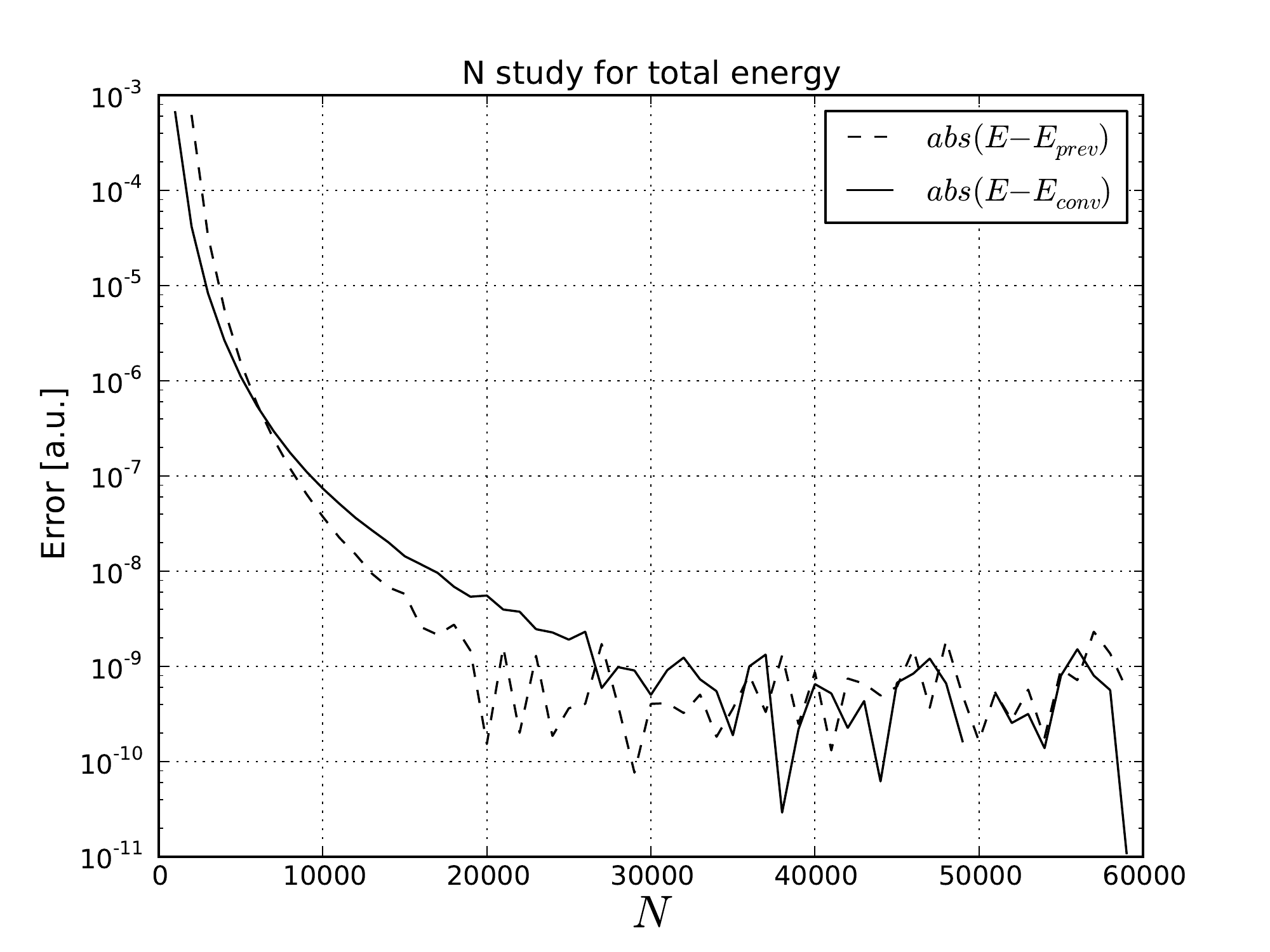}
}
\caption{Relativistic (RLDA) density functional theory (DFT) convergence
studies for uranium.}
\end{figure}

We first determine sufficiently large $r_\max$ that the associated error (due to confining wavefunctions and potentials) is below the level of other sources in the calculation. We start by setting $r_\min$, $r_\max$, $a$, and $N$ to typical values consistent with previous findings~\cite{nist}. We then increase $N$ to converge to the limit of finite precision. 
Then, increasing $r_\max$ in LDA (Figs.~\ref{fig:lda_rmax_eig} and
\ref{fig:lda_rmax_etot}) and RLDA (Figs.~\ref{fig:rlda_rmax_eig} and
\ref{fig:rlda_rmax_etot}) calculations (with perturbation correction turned off), 
considering both eigenvalues and total energy, shows that $r_\max=50\rm\,a.u.$ is a
well converged value for both LDA and RLDA, consistent with previous findings~\cite{nist}.

The aim of the next $N$ study is then to find $N$ (and $a$) such that the total
energy is converged for broad range of $r_\min$, so that such $N$ can
be used for the $r_\min$ study. In theory, by simply increasing $N$, the total energy must eventually
converge, but for small $a$ (low concentration of grid points around the origin), 
the required $N$ might be very high.
There is some coupling of $r_\min$ and $a$ due to closer approach to the singularity 
at $r=0$ for smaller $r_\min$: smaller $r_\min$ requires somewhat larger $a$ for a given $N$ and accuracy. It was found that using $a=10^9$, full convergence is achieved for
all $r_\min > 10^{-14}$ with $N = 50000$. 

The initial point $r_\min>0$ determines the accuracy of the asymptotics used to
start the radial integrations, which are exact only in the limit $r \to 0$.
The goal of the $r_\min$ study is to find fully converged $r_\min$ such that
the associated error is at the limit of finite precision;  so that decreasing
further does not improve results.  To do so, the fully converged $a$ and $N$
are taken from the initial $N$ study, then converged total energies are
calculated for all $r_\min$ from $10^{-14}$ to $10^{-5}$.

Fig.~\ref{fig:lda_rmin} shows the $r_\min$ study for the LDA calculation. 
As can be seen, the total energy is converged for $r_\min = 10^{-7}$
and decreasing $r_\min$ further, the error indicators remain below the
$5\cdot 10^{-9}$ level.
Choosing $r_\min$ smaller than $10^{-7}$ does not increase accuracy further.
In general, choosing $r_\min$ as large as possible avoids the need for
excessive grid points to resolve rapidly varying potentials, 
densities, and wavefunctions in the vicinity of the Coulomb singularity at $r=0$.
As such, $r_\min = 10^{-7}$ is used, as
it gives sufficiently accurate asymptotics such that the associated errors are reduced to the limits of finite precision
(which eliminates $r_\min$ from consideration when choosing other parameters), is consistent with values adopted previously~\cite{nist},
and is not so small as to require excessive grid points around the origin.

Fig.~\ref{fig:rlda_rmin} shows the $r_\min$ study for the RLDA calculation.
Using the same procedure as for the LDA calculation, $r_\min =
10^{-8}$ was chosen.

Now the optimal mesh for $10^{-6}$~a.u. accuracy in total energy 
is found by setting $r_\min$ and $r_\max$ to converged values 
and varying the mesh gradation $a$ to minimize the number of elements $N$ required to achieve the specified accuracy. 
Proceeding in this way, we find optimal mesh parameters as in Table~\ref{table:meshreference} 
for $10^{-6}$~a.u. accuracy in the total energy of uranium.

Having determined converged $r_\min$, $r_\max$, and optimized $a$
(Table~\ref{table:meshreference}), 
we take $N$ to convergence (Fig.~\ref{fig:lda_N}) to find converged total energy
\begin{equation}
E_{tot}= -25658.41788885\rm\,a.u.
\end{equation}
and orbital energies (Table~\ref{table:U_eigs})
with error $<10^{-8}\rm\,a.u.$ for the LDA calculation.

\begin{table}
\begin{center}
{\tiny
\begin{tabular}{c r}
state & eigenvalue [a.u.]\\
\hline
1s &-3689.35513984 \\
2s & -639.77872809 \\
2p & -619.10855018 \\
3s & -161.11807321 \\
3p & -150.97898016 \\
3d & -131.97735828 \\
4s &  -40.52808425 \\
4p &  -35.85332083 \\
4d &  -27.12321230 \\
4f &  -15.02746007 \\
5s &   -8.82408940 \\
5p &   -7.01809220 \\
5d &   -3.86617513 \\
5f &   -0.36654335 \\
6s &   -1.32597632 \\
6p &   -0.82253797 \\
6d &   -0.14319018 \\
7s &   -0.13094786 \\
\end{tabular}
}
\caption{Computed non-relativistic (LDA) eigenvalues for uranium ($10^{-8}\rm\,a.u.$ accurate).}
\label{table:U_eigs}
\end{center}
\end{table}

Proceeding similarly for the RLDA calculation, 
we take $N$ to convergence (Fig.~\ref{fig:rlda_N}) to find converged total energy 
\begin{equation}
E_{tot}= -28001.13232548\rm\,a.u.
\end{equation}
and orbital energies (Table~\ref{table:U_eigs_rel}) 
with error $<10^{-8}\rm\,a.u.$ for the RLDA calculation also.

\begin{table}
\begin{center}
{\tiny
\begin{tabular}{c r}
state & eigenvalue [a.u.] \\
\hline
$1s_{1/2}$ & -4223.41902045 \\
$2s_{1/2}$ & -789.48978233 \\
$2p_{3/2}$ & -761.37447597 \\
$2p_{1/2}$ & -622.84809456 \\
$3s_{1/2}$ & -199.42980564 \\
$3p_{3/2}$ & -186.66371312 \\
$3p_{1/2}$ & -154.70102667 \\
$3d_{5/2}$ & -134.54118029 \\
$3d_{3/2}$ & -128.01665738 \\
$4s_{1/2}$ & -50.78894806 \\
$4p_{3/2}$ & -45.03717129 \\
$4p_{1/2}$ & -36.68861049 \\
$4d_{5/2}$ & -27.52930624 \\
$4d_{3/2}$ & -25.98542891 \\
$4f_{7/2}$ & -13.88951423 \\
$4f_{5/2}$ & -13.48546969 \\
$5s_{1/2}$ & -11.29558710 \\
$5p_{3/2}$ & -9.05796425 \\
$5p_{1/2}$ & -7.06929564 \\
$5d_{5/2}$ & -3.79741623 \\
$5d_{3/2}$ & -3.50121718 \\
$5f_{7/2}$ & -0.14678839 \\
$5f_{5/2}$ & -0.11604717 \\
$6s_{1/2}$ & -1.74803995 \\
$6p_{3/2}$ & -1.10111900 \\
$6p_{1/2}$ & -0.77578418 \\
$6d_{5/2}$ & -0.10304082 \\
$6d_{3/2}$ & -0.08480202 \\
$7s_{1/2}$ & -0.16094728 \\
\end{tabular}
}
\caption{Computed relativistic (RLDA) eigenvalues for uranium ($10^{-8}\rm\,a.u.$ accurate).}
\label{table:U_eigs_rel}
\end{center}
\end{table}

\subsubsection{Optimal $N$}

In the previous section, for converged $r_\min$, $r_\max$, and optimized $a$
(Table~\ref{table:meshreference}), 
sufficient $N$ was determined to achieve $10^{-6}$ accuracy in total energies for uranium.
In this section, we determine sufficient $N$ for a series of accuracies, from $10^{-3}$ to $10^{-8}$~a.u., for all elements $Z = 1$ -- 92.

\begin{figure}
\centering
\subfloat[Schr\"odinger (LDA)]{
\label{fig:acc_schroed}
\includegraphics[width=0.5\linewidth]{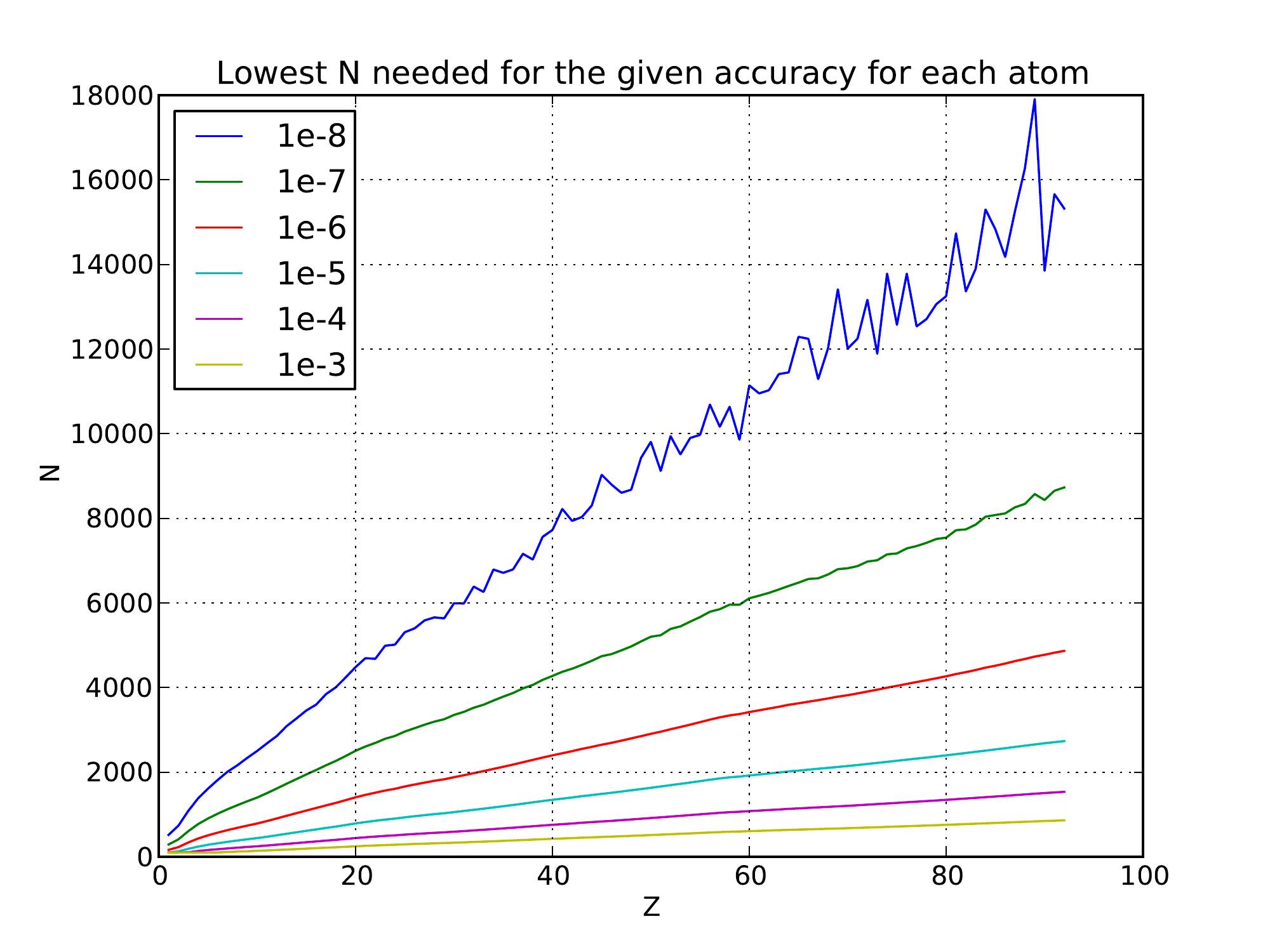}
}
\subfloat[Dirac (RLDA)]{
\label{fig:acc_dirac}
\includegraphics[width=0.5\linewidth]{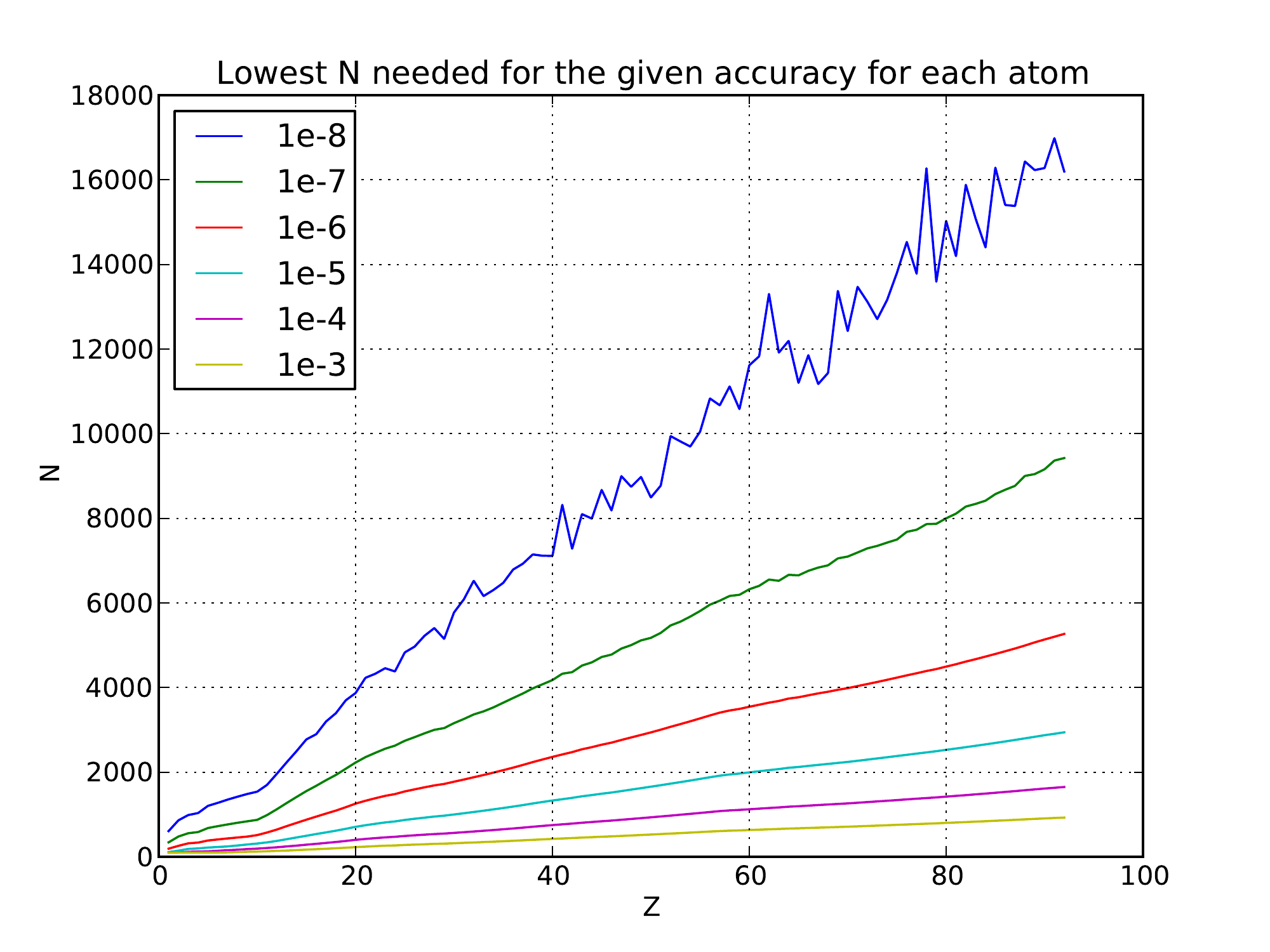}
}
\caption{Accuracy graphs for non-relativistic (LDA) and relativistic (RLDA)
density functional theory (DFT) calculations.}
\end{figure}

Fig.~\ref{fig:acc_schroed} shows the required $N$ for LDA (Schr\"odinger) and 
Fig.~\ref{fig:acc_dirac} for RLDA (Dirac) Kohn--Sham equations.

For simplicity, we use the same converged $r_\min$, $r_\max$, and optimized $a$ 
(Table~\ref{table:meshreference})
for all atoms and just vary $N$ for each atom to find a sufficient value for the specified accuracy.

The $N$ shown in Figs.~\ref{fig:acc_schroed} and \ref{fig:acc_dirac} is mainly
for use as a starting point. Sufficient $N$ will vary somewhat with the choice of
exchange-correlation, and very much so with the choice of $r_\min$, $r_\max$, and
$a$. If any of these are changed, as may be desired to find more optimal values
for different atoms and/or accuracy requirements, a new $N$ study will be
required to determine sufficient $N$. If, however, the converged $r_\min$,
$r_\max$, and $a$ from Table~\ref{table:meshreference} 
are used, then convergence can be attained for any atom to the limits of machine precision by simply increasing $N$.

By setting $N=50000$ with converged $r_\min$, $r_\max$, and $a$ 
from Table~\ref{table:meshreference}, 
fully converged total energies and Kohn--Sham eigenvalues are obtained, 
and used to compute errors for 
Figs.~\ref{fig:acc_schroed} and~\ref{fig:acc_dirac}.

\subsubsection{Numerical error}

\begin{figure}
\centering
\subfloat[Schr\"odinger]{
\label{fig:lda_N_log}
\includegraphics[width=0.5\linewidth]{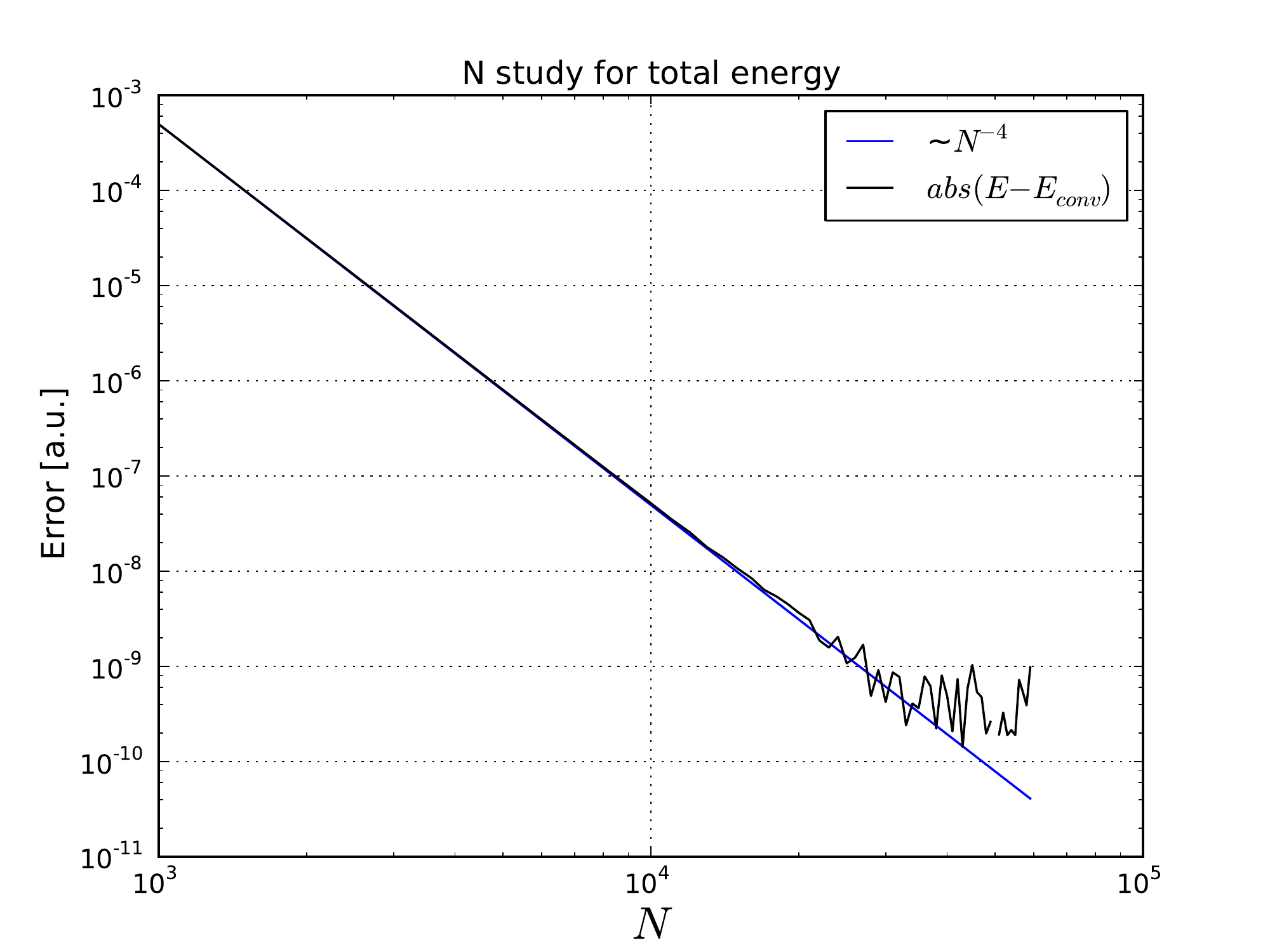}
}
\subfloat[Dirac]{
\label{fig:rlda_N_log}
\includegraphics[width=0.5\linewidth]{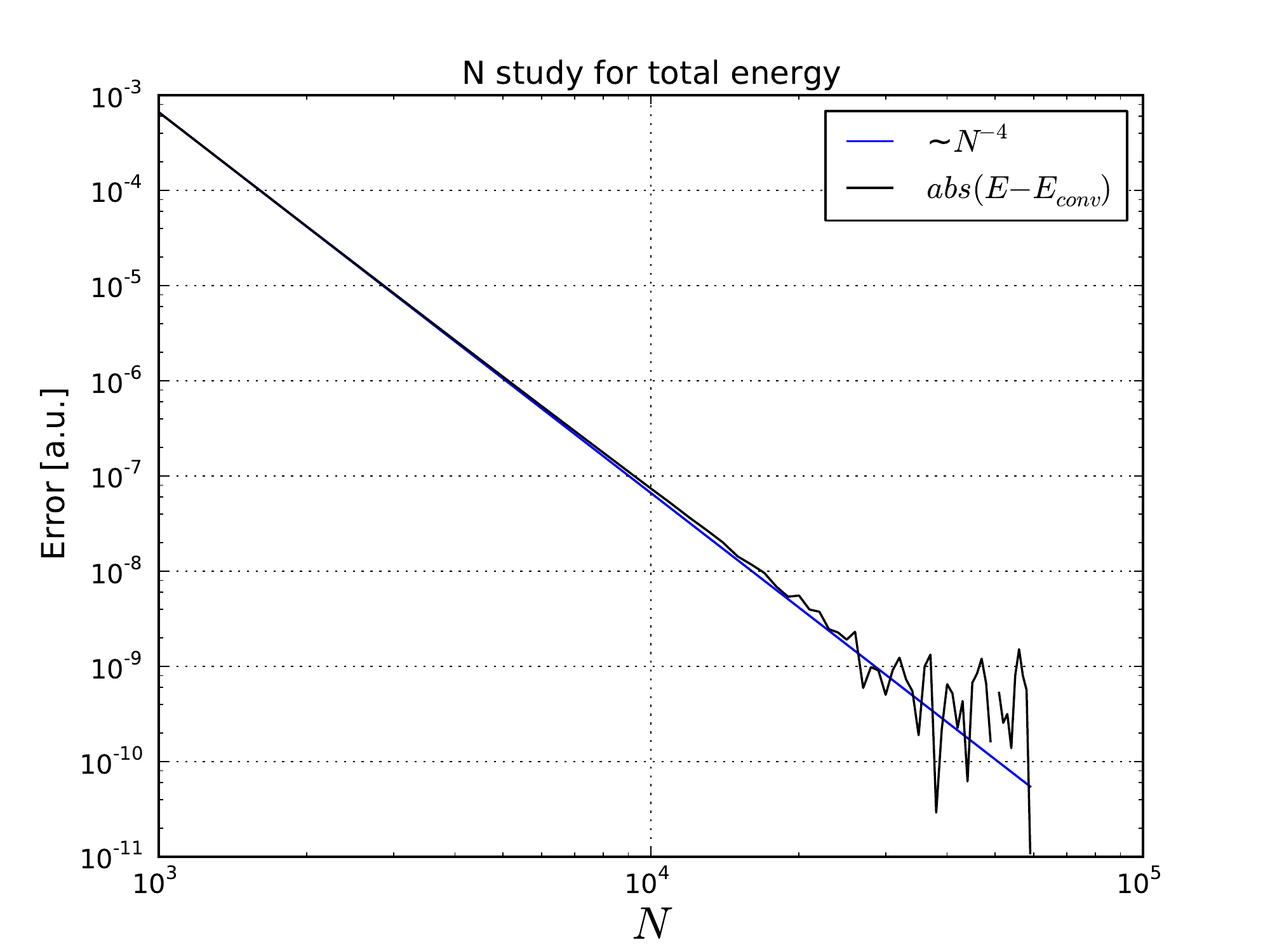}
}
\caption{Non-relativistic (LDA) and relativistic (RLDA) convergence rates from
the $N$-studies for total energy plotted on a log-log scale.}
\end{figure}

The intrinsic numerical error of the Schr\"odinger, Dirac, and Poisson solvers
is $O(h^4)$, where $h$ is the mesh spacing, since we use 
4th-order Runge-Kutta and Adams methods. For the uranium test case, 
the convergence rate in terms of the number of mesh points $N$ ($\propto 1/h$) can be
estimated from the $N$-convergence studies (Fig.~\ref{fig:lda_N} and
\ref{fig:rlda_N}) plotted on a log-log scale, as shown in Figs.~\ref{fig:lda_N_log}
and \ref{fig:rlda_N_log}. For comparison, the line $N^{-4}$ is plotted also (shifted to agree at lowest $N$). As
can be seen, the actual convergence follows $N^{-4}$ almost exactly, indicating an error of $O(h^4)$, consistent with the order of methods employed.

\subsubsection{Comparison to current benchmarks}

Using the meshes from 
Table~\ref{table:meshreference} 
with converged $N=50000$, the computed Kohn--Sham total energies 
and eigenvalues were compared with current benchmark values~\cite{nist,nistweb} 
for all atoms $Z = 1$ -- 92 
for LDA (Schr\"odinger) and RLDA (Dirac) Kohn--Sham calculations.
It was found that the computed total energies were all within $5.29 \cdot 10^{-7}$
of benchmark values for LDA and within $5.31 \cdot 10^{-7}$ for RLDA;  
and that the computed eigenvalues were all within $1.24\times 10^{-6}$ 
of benchmark values for LDA and within $1.37\times 10^{-6}$ for RLDA.

Reference~\cite{nist} claims $10^{-6}$~a.u. absolute accuracy for Kohn--Sham 
total energies and finds
maximum deviations among different codes of $2 \cdot 10^{-6}$ 
for associated eigenvalues. Our results, converged to $<10^{-8}$ a.u., verify that the total energies reported in \cite{nist} 
indeed have absolute accuracy $10^{-6}$~a.u. or better; and furthermore, that the eigenvalues in \cite{nist} in fact have \tit{absolute} accuracy $2 \cdot 10^{-6}$~a.u. or better.

\section{Summary and conclusions}

\label{conclusions}

We have presented a robust and general solver for the solution of the radial
Schr\"odinger, Dirac, and Kohn--Sham equations of density functional theory;
and provided a modular, portable, and efficient Fortran 95 implementation,
\ttt{dftatom}, along with interfaces to other languages and full suite of
examples and tests. The solver brings together ideas from many different codes
developed over the decades and in addition 
employs ideas such as outward Poisson
integration for increased accuracy in the core region and perturbation with
fallback to bisection for speed and robustness. The solver can accommodate any
potential, whether singular Coulombic or finite, and any mesh, whether linear,
exponential, or otherwise. We have demonstrated the flexibility and accuracy of
the associated code with solutions of Schr\"odinger and Dirac equations for
Coulombic, harmonic oscillator, and double minimum potentials; 
and solutions of Kohn--Sham and
Dirac--Kohn--Sham equations for the challenging case of uranium, obtaining
energies accurate to $10^{-8}$ a.u., thus verifying and refining by two orders
of magnitude current benchmarks \cite{nist}. We have shown detailed convergence
studies in each case, providing mesh parameters to facilitate straightforward
convergence to any desired accuracy by simply increasing the number of mesh
points.

At all points in the design of the associated code, we have tried to emphasize simplicity and modularity so that the routines provided can be straightforwardly employed for a range of applications purposes, while retaining high efficiency. We have made the code available as open source to facilitate distribution, modification, and use as needed. We expect the present solvers will be of benefit to a range of larger-scale electronic structure methods which rely on atomic structure calculations and/or radial integrations more generally as key components.

\section{Acknowledgements}

We would like to thank Don Hamann and Mark Stiles for discussing and sharing
their codes with us, as well as Eric Shirley for pointing us to other codes. We
thank Charlotte Froese Fischer and Zbigniew Romanowski for helpful discussions
and Peter Winkler for a critical reading of the manuscript. Finally, we 
 thank the referees for a number of valuable suggestions which improved the manuscript considerably.

This work performed, in part, under the auspices of the U.S. Department of
Energy by Lawrence Livermore National Laboratory under Contract
DE-AC52-07NA27344. This research was partly supported by the LC06040 research
center project and GACR 101/09/1630 of the Czech Science Foundation.

\bibliographystyle{cpc}
\bibliography{certik}

\end{document}